\documentclass[aps,prc,twocolumn,nofootinbib,amsmath,showpacs,superscriptaddress,floatfix]{revtex4-1}
\usepackage{graphicx,epsfig,epstopdf,longtable}
\usepackage{mathptmx}
\usepackage{xcolor}
\usepackage[light,first]{draftcopy} 
\usepackage[pdftex]{hyperref}
\usepackage[mathlines]{lineno}
\usepackage{wasysym}
\begin{document}

\preprint{APS/123-QED}

\title{Verification of detailed balance for $\gamma$ absorption and emission in Dy isotopes}

\author{T.~Renstr{\o}m}
\email{therese.renstrom@fys.uio.no}
\affiliation{Department of Physics, University of Oslo, N-0316 Oslo, Norway}

\author{H.~Utsunomiya}
\affiliation{Department of Physics, Konan University, Okamoto 8-9-1, Higashinada, Kobe 658-8501, Japan}

\author{H.~T.~Nyhus}
\affiliation{Department of Physics, University of Oslo, N-0316 Oslo, Norway}

\author{A.~C.~Larsen}
\affiliation{Department of Physics, University of Oslo, N-0316 Oslo, Norway}

\author{M.~Guttormsen}
\affiliation{Department of Physics, University of Oslo, N-0316 Oslo, Norway}

\author{G.~M.~Tveten}
\affiliation{Department of Physics, University of Oslo, N-0316 Oslo, Norway}

\author{D.~M.~Filipescu}
\affiliation{Extreme Light Infrastructure Nuclear Physics, "Horia Hulubei" National
Institute for Physics and Nuclear Engineering (IFIN-HH), 30 Reactorului,
077125 Bucharest-Magurele, Romania}
\affiliation{"Horia Hulubei" National Institute for Physics and Nuclear Engineering
(IFIN-HH), 30 Reactorului, 077125 Bucharest-Magurele, Romania}

\author{I.~Gheorghe}
\affiliation{Extreme Light Infrastructure Nuclear Physics, "Horia Hulubei" National
Institute for Physics and Nuclear Engineering (IFIN-HH), 30 Reactorului,
077125 Bucharest-Magurele, Romania}
\affiliation{"Horia Hulubei" National Institute for Physics and Nuclear Engineering
(IFIN-HH), 30 Reactorului, 077125 Bucharest-Magurele, Romania}

\author{S.~Goriely}
\affiliation{Institut d'Astronomie et d'Astrophysique, Universit\'{e} Libre de Bruxelles, Campus de la Plaine, CP-226, 1050 Brussels, Belgium}

\author{S.~Hilaire} 
\affiliation{CEA, DAM, DIF, F-91297 Arpajon, France}

\author{Y.-W.~Lui}
\affiliation{Cyclotron Institute, Texas A\& M University, College Station, Texas 77843, USA}

\author{J.~E.~Midtb{\o}}
\affiliation{Department of Physics, University of Oslo, N-0316 Oslo, Norway}

\author{S.~P\'{e}ru}
\affiliation{CEA, DAM, DIF, F-91297 Arpajon, France}

\author{T.~Shima}
\affiliation{Research Center for Nuclear Physics, Osaka University, Suita, Osaka 567-0047, Japan}

\author{S.~Siem}
\affiliation{Department of Physics, University of Oslo, N-0316 Oslo, Norway}

\author{O.~Tesileanu}
\affiliation{Extreme Light Infrastructure Nuclear Physics, str Atomistilor nr. 407, Bucharest-Magurele, P.O.BOX MG6, Romania}

\date{\today}

\begin{abstract}
The photo-neutron cross sections of $^{162,163}\rm{Dy}$ have been measured
for the first time in an energy region from the neutron threshold ($S_n$) up to $\approx$ $13$~MeV.
The ($\gamma$,n) reaction was induced with quasi-monochromatic laser
Compton-scattered $\gamma$ rays, produced at the NewSUBARU laboratory.
The corresponding $\gamma$-ray strength functions ($\gamma$SF) have been calculated from the
photo-neutron cross sections. The data are compared to reanalyzed $\gamma$SFs
of $^{160-164}\rm{Dy}$, which are measured below $S_n$. 
The excellent agreement with the photo-neutron data at $S_n$ confirms the principle of detailed balance. 
Thus, a complete $\gamma$SF is established covering in total the energy region of 1 MeV $\leq$  E$_{\gamma}$ $\leq$ 13 MeV.
These mid-shell well-deformed dysprosium isotopes all show scissors resonances with very similar structures. We find that our data predict the same integrated scissors strength as ($\gamma,\gamma^\prime$) data when integrated over the same energy range, which shows that the scissors mode very likely is consistent with the generalized Brink hypothesis. 
Finally, using the $\gamma$SFs as input in the reaction code TALYS, we have deduced radiative neutron-capture cross sections and compared them to direct measurements. 
We find a very good agreement within the
uncertainties, which gives further support to the experimentally determined $\gamma$SFs. 
\end{abstract}

\maketitle

\section{Introduction}
The principle of detailed balance is 
one of the most fundamental assumptions
commonly used in quantum mechanics. 
The background for this principle is the
fact that inverse processes are strongly 
linked~\cite{blatt_weisskopf}. 
As stated by Blatt and Weisskopf, 
this principle
\textit{''can be applied ... to the emission 
and absorption of $\gamma$ radiation in nuclei''}. 
In nuclear physics,
detailed balance is often used in the description of electric dipole ($E1$) absorption and emission. 
For example,
state-of-the-art microscopic calculations of ground-state $E1$ excitations~\cite{martini2015}, probing the giant electric dipole resonance (GEDR), assume detailed balance to estimate radiative neutron-capture cross sections for nuclear-astrophysics applications. 

Moreover, the Brink hypothesis~\cite{brink1955}
states that \textit{''... if it were possible to perform the photoeffect on an excited state, the cross section for absorption would have the same energy dependence as for the ground state''}. This hypothesis is used together with the principle of detailed balance
to calculate average, total radiative widths and radiative neutron-capture cross sections.

However, to verify the application of detailed balance and the Brink hypothesis to obtain a complete description of $\gamma$ absorption and decay, 
one needs to measure both photo-neutron data above the neutron threshold $S_n \approx 6-8.5$ MeV,
and $\gamma$-decay data below $S_n$.  
If these data sets agree with each other, 
it is a strong indication that detailed balance and the Brink hypothesis are indeed fulfilled. 

For the GEDR, the Brink hypothesis seems to be valid except for nuclear reactions involving high temperatures and/or spins~\cite{andreas_michael}. 
However, this is not necessarily true for other
types of $\gamma$ resonances.
At the low-energy tail of the GEDR, other resonance  
structures appear as well, such as the pygmy dipole resonance (PDR)~\cite{savran2013}, the magnetic-dipole ($M1$) spin-flip resonance~\cite{KHeyde_magnetisk_dipol_review}, 
and the $M1$ scissors resonance (SR) built on the ground state~\cite{KHeyde_magnetisk_dipol_review}
and on excited states~\cite{Krticka_letter,schiller2006}. 

The first experimental indication of the SR on excited states (quasi-continuum region), was the observation of a strong enhancement in the $\gamma$ spectrum of excited $^{161}$Dy at around 3 MeV in 1984~\cite{Magne_scissor_1984}. 
This structure was interpreted as an implication of the SR predicted to occur in deformed nuclei~\cite{Lo_first_scissor_prediction,Leander_first_nilsson_scissor}. 
Later in the same year, the (e,e$^{\prime}$) reaction was used to reveal $M1$ type resonant states in Gd~\cite{First_e_e} built on the ground state. 

For rare-earth Dy isotopes, the integrated SR strength as reported from Nuclear Resonance Fluorescence (NRF) measurements~\cite{Kneissl} is about half the summed strength found in Oslo-type experiments~\cite{Magne2003,Hilde2010} and by two-step cascade (TSC) measurements following thermal neutron capture~\cite{Krticka_letter}.
It has been suggested~\cite{guttormsen2014} that the discrepancy could be due to differences in the nuclear moment of inertia, as the summed $B_{\rm SR}$ strength is predicted to be proportional with the SR moment of inertia~\cite{Iudice1993}.
As the ground-state moment of inertia is smaller than the moment of inertia for excited states~\cite{guttormsen2014}, this could possibly explain the observed discrepancy. 
On the other hand, this explanation would be in conflict with the generalized Brink hypothesis, which is understood in the following way: 
any collective mode has the same properties regardless of whether it is built on the ground state or excited states. 
This is, however, still an open question, because results from Ref.~\cite{Guttormsen2016} provide evidence for the generalized Brink hypothesis also for transitions to the ground band in $^{238}$Np.

Another explanation for the observed deviations could be the fact that disentangling the SR strength from different types of backgrounds is a challenge in all the aforementioned experimental methods. 
For example, in the Oslo-type and TSC experiments, it is necessary to assume an underlying $E1$ "tail", which must be subtracted to  estimate the SR strength, because the $M1$ and $E1$ components cannot be distinguished from the data.
For NRF experiments, weaker transitions might be hidden in the atomic background, while the low intensity of the endpoint-bremsstrahlung spectrum could make it difficult to measure transitions near the endpoint energy. 
Further, NRF data on the SR strength have typically been evaluated at a narrow excitation-energy region ($\approx 2.7 -3.7$ MeV), while Oslo-type and TSC experiments give a summed strength for a large energy interval. 

As photo-neutron data provide a measure on the GEDR and hence the dominant $E1$ strength, a good determination of this component is crucial for extracting the summed SR strength in Oslo-type experiments, which cannot separate $E1$ and $M1$ radiation directly. 
Surprisingly, the otherwise well-studied rare-earth nuclei are relatively unexplored in the energy region above $S_n$. 

Of the 33 stable well-deformed rare-earth nuclei from $^{154}$Sm to $^{176}$Lu ($\beta_2 >0.3$~\cite{def}), there exists photo-neutron data on only six nuclei~\cite{EXFOR}.  

In this work, we report for the first time on photo-neutron measurements of $^{162,163}$Dy, ranging from excitation energies of $S_n$ and up to $\approx$ 13 MeV. 
With these new measurements, we address two main questions:
(\textit{i}) Is the principle of detailed balance fulfilled? 
(\textit{ii}) Is the generalized Brink hypothesis valid for the SR?
In an attempt to answer these questions, the photo-neutron data will be combined with the reanalyzed $\gamma$SFs below $S_n$ of the $^{160-164}$Dy isotopes using the Oslo method. 

Furthermore, we reevaluate the summed strengths and uncertainties of the SRs and compare with TSC and NRF data, in addition to new results from multistep-cascade (MSC) measurements of $\gamma$ decay following neutron capture from a white neutron source~\cite{valenta2017}. 
Finally, on the basis of the reanalyzed NLDs and $\gamma$SFs, we calculate radiative neutron-capture cross sections within the Hauser-Feshbach formalism and compare with experimental ($n,\gamma$) data. 
This is the final test of the two questions raised above: if the principle of detailed balance and the generalized Brink hypothesis are applicable, one would expect a good reproduction of direct ($n,\gamma$) cross-section measurements.

\section{Experimental procedure} 
\label{sec_exp}
The photo-neutron measurements on $^{162,163}$Dy took place at the NewSUBARU synchrotronic radiation facility~\cite{NewSUB_website}. Here, quasi-monochromatic $\gamma$-ray beams were produced through laser
Compton scattering (LCS) of 1064 nm photons in head-on collisions with relativistic electrons. Throughout the experiment, the laser was periodically on for 80 ms and off for 20 ms, in order to measure background neutrons and $\gamma$-rays. The electrons were injected from a linear accelerator into the NewSUBARU storage ring~\cite{exp} with an initial
energy of $\approx 1$~GeV, then subsequently deaccelerated to energies in the region from $\approx600$~MeV to
$\approx900$~MeV, providing LCS $\gamma$-ray beams
from $S_n$ up to $E_{\gamma}\approx 13$~MeV. In total, 12 individual $\gamma$ beams were produced  for $^{162}$Dy and 15 for $^{163}$Dy. The energy profiles of the produced $\gamma$-ray beams were measured
with a $3.5^{\prime\prime}\times 4.0^{\prime\prime}$ LaBr$_3$:Ce (LaBr$_3$) detector. The measured LaBr$_3$ spectra
were reproduced by a GEANT4 code~\cite{Ioana_thesis}, which takes into account the kinematics of the LCS process, including the beam emittance and the interactions between the LCS beam and the LaBr$_3$ detector. In this way we were able to simulate the incoming energy profile of the $\gamma$ beams.

The $^{162,163}$Dy targets were in oxide form with an areal density of $2.21$~g/cm$^2$ and $1.94$~g/cm$^2$, respectively.
The corresponding enrichments of the two isotopes were $99.28\%$ and $96.85\%$. The target material was placed inside aluminum containers. For neutron detection, a high-efficiency $4\pi$ detector was used, consisting of 20 $^3\rm{He}$ proportional counters, arranged in three concentric rings and embedded in a 36 $\times$ 36 $\times$ 50 cm$^3$ polyethylene neutron moderator. The average energy of the neutrons from the ($\gamma$,n) reactions were estimated using the ring ratio technique, originally developed by Berman~\cite{Berman_ring_ratio}. The efficiency of the neutron detector depends on the neutron energy, and was found by Monte Carlo simulations. Thereby we are able to deduce the number of ($\gamma$, n) reactions that took place during each run.

\begin{figure}
\includegraphics[totalheight=4.0cm,angle=0]{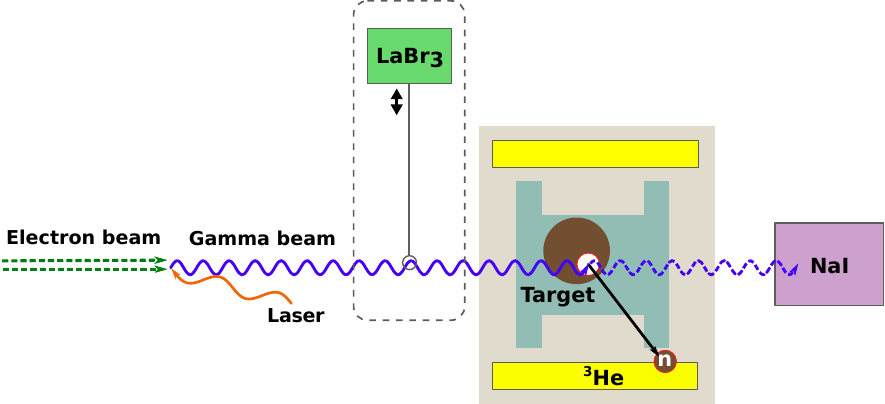}
\caption{(Color online) A schematic illustration of the experimental set up at NewSUBARU.}
\label{fig:setup}
\end{figure}

The LCS $\gamma$-ray flux was monitored by a $5^{\prime\prime}\times 6^{\prime\prime}$ NaI:Tl (NaI)
detector during neutron measurement runs. The number of incoming $\gamma$ rays per measurement was estimated using the pile-up technique described in Ref.~\cite{kondo}.
Figure~\ref{fig:setup} shows a schematic illustration of the experimental set up.

The measured photo-neutron cross section for an incoming beam with maximum $\gamma$-energy $E_{\rm max}$ is given by the convoluted cross section,
\begin{equation}
\sigma^{E_{\rm max}}_{\rm exp}=\int_{S_n}^{E_{\rm max}}D^{E_{\rm max}}(E_{\gamma})\sigma(E_{\gamma})dE_{\gamma}=\frac{N_n}{N_tN_{\gamma}\xi\epsilon_n g}.
\label{eq:cross1}
\end{equation}
Here, $D^{E_{\rm max}}$ is the normalized,$\int_{S_n}^{E_{\rm max}} D^{E_{\rm max}}dE_{\gamma}= 1$, energy distribution of
the $\gamma$-ray beam obtained from GEANT4 simulations. The simulated profiles of the $\gamma$ beams, $D^{E_{\rm max}}$, used to investigate $^{163}$Dy are shown in Fig.~\ref{fig:GammaProfile}.  Furthermore, $\sigma(E_{\gamma})$ is the true
photo-neutron cross section as a function of energy. The quantity $N_n$ represents the number of
neutrons detected, $N_t$ gives the number of target nuclei per unit area, $N_{\gamma}$
is the number of $\gamma$ rays incident on target, $\epsilon_n$ represents the
neutron detection efficiency, and finally $\xi=(1-e^{\mu t})/(\mu t)$ gives a
correction factor for self-attenuation in the target. The factor $g$ represents
the fraction of the $\gamma$ flux above $S_n$. 

\begin{figure}[]
\includegraphics[clip,width=1.\columnwidth]{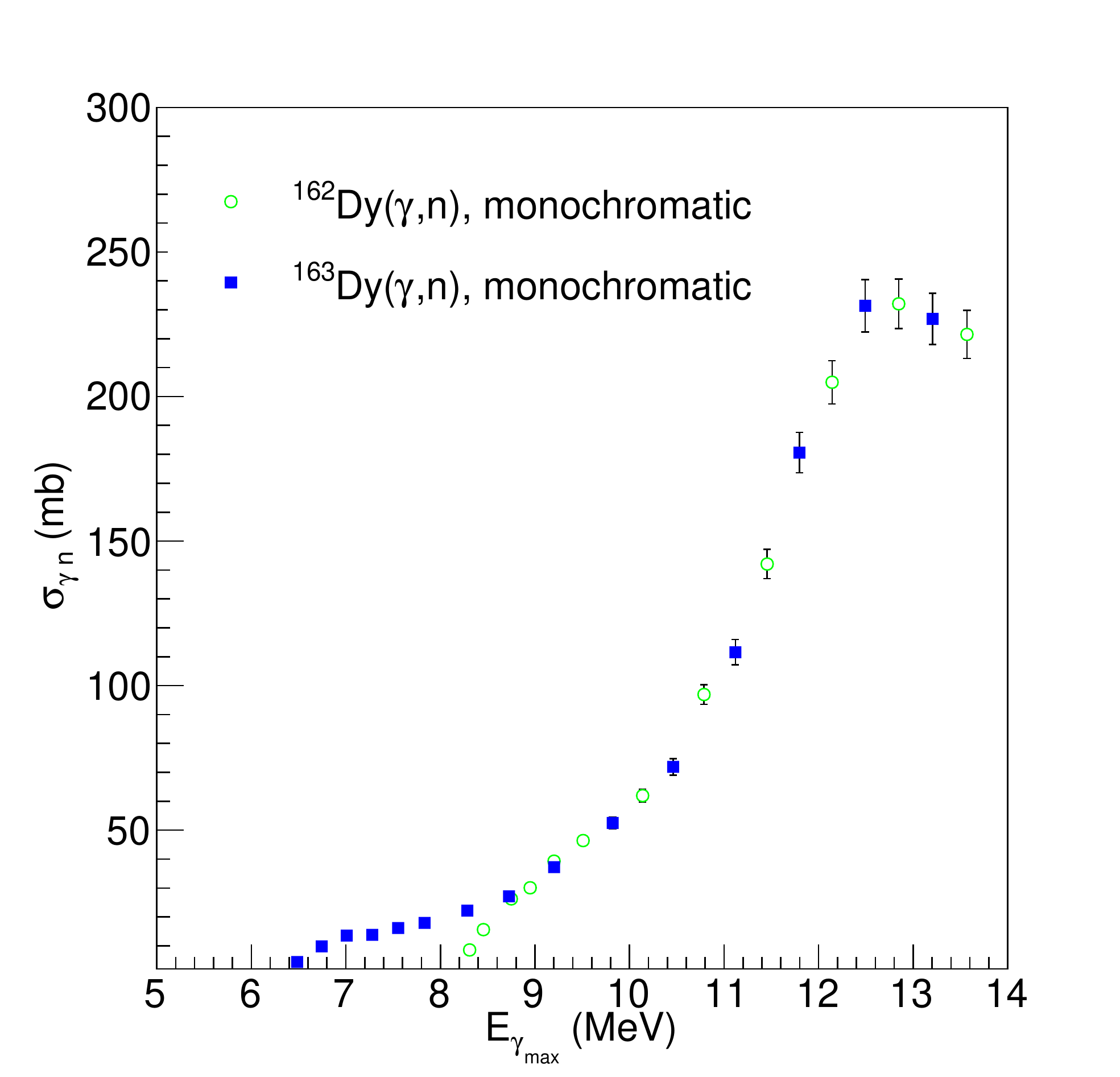}
\caption{(Color online) Monochromatic cross sections of $^{162}\rm{Dy}$ (green open circles) and $^{163}\rm{Dy}$ (blue filled squares). The error bars contain statistical uncertainties from the number of detected neutrons, the uncertainty in the efficiency of the neutron detector and the uncertainly in the pile-up method used to determine the number of incoming $\gamma$'s on target.}
\label{fig:MonocrossBoth}
\end{figure}

We have determined the convoluted cross sections $\sigma^{E_{\rm max}}_{\rm exp}$ given by Eq.~(\ref{eq:cross1}) for $\gamma$ beams with maximum energies in the range $S_{n}\leq E_{\rm max} \leq$ 13 MeV. The convoluted cross sections $\sigma^{E_{\rm max}}_{\rm exp}$ are not connected to a specific $E_{\gamma}$, and we choose to plot them as a function of $E_{\gamma \rm max}$. The convoluted cross sections of the two Dy isotopes, which are often called monochromatic cross sections, are shown in Fig.~\ref{fig:MonocrossBoth}. The error bars in Fig.~\ref{fig:MonocrossBoth} represent the total uncertainty in the quantities comprising Eq.~(\ref{eq:cross1}) and consists of $\sim 3.2\%$ from the efficiency of the neutron detector, $\sim 3\%$ from the pile-up method that gives the number of $\gamma$-rays, and the statistical uncertainty in the number of detected neutrons.
For the total uncertainty, we have assumed that the errors are independent and thus added quadratically.

\section{Data analysis}
\label{sec_data}
The challenge now is to extract the deconvoluted, $E_{\gamma}$ dependent, photo-neutron cross section, $\sigma(E_{\gamma})$, from the integral of Eq.~(\ref{eq:cross1}).
Each of the measurements characterized by the beam energy, $E_{\rm max}$, correspond to folding of $\sigma(E_{\gamma})$ with the 
measured beam profile, $D^{E_{\rm max}}$.  
\begin{figure}[]
\includegraphics[clip,height=8cm]{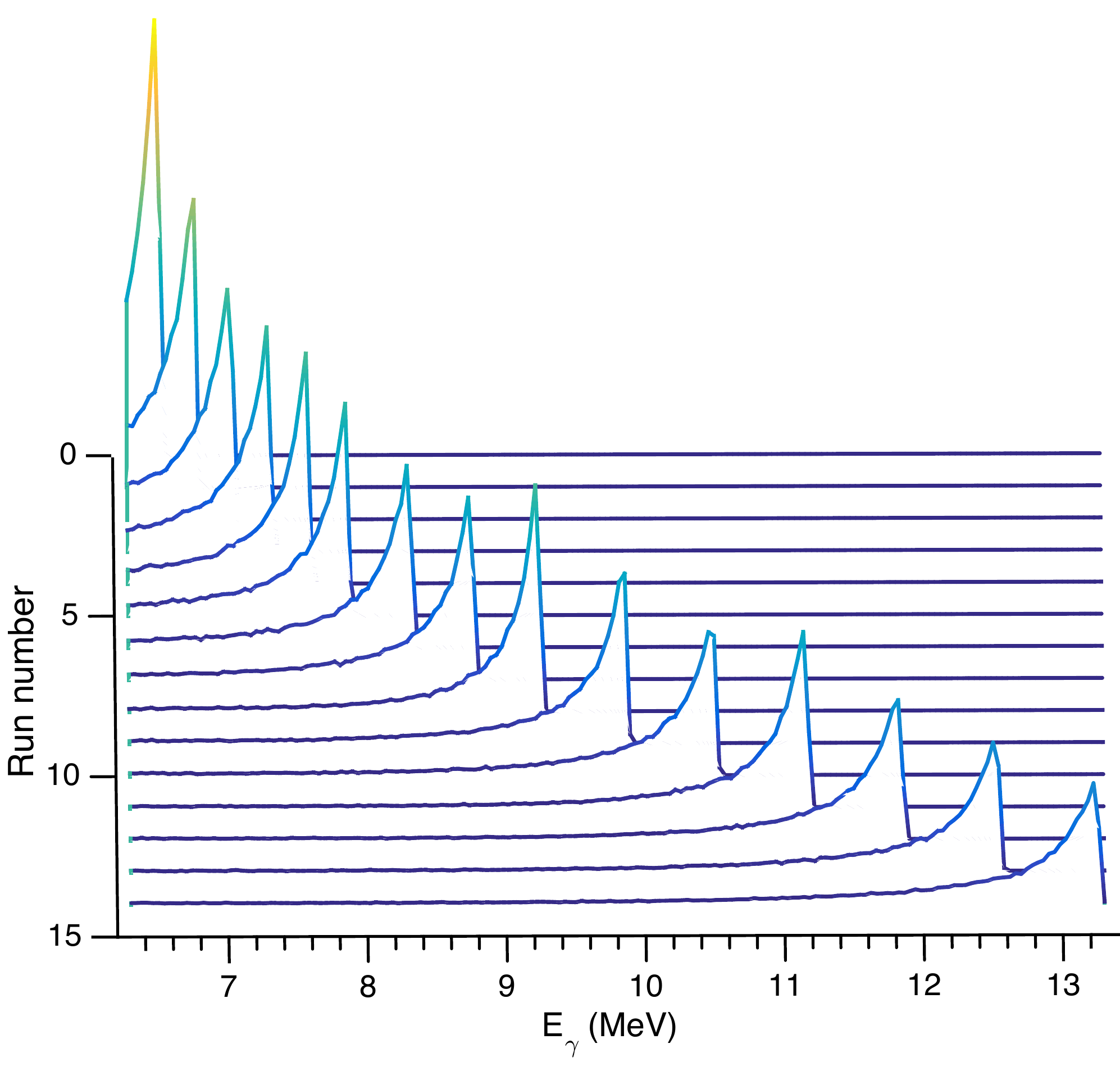}
\caption{(Color online) The simulated $\gamma$-beam profiles in the case of $^{163}$Dy.}
\label{fig:GammaProfile}
\end{figure}
By approximating the integral in Eq.~(\ref{eq:cross1}) with a sum for each $\gamma$-beam profile, we are able to express the problem as a set of linear equations
\begin{equation}
\sigma_{\rm f }=\bf{D}\sigma,
\end{equation}
where $\sigma_{\rm f}$ is the cross section folded with the beam profile {\bf D}.  
The indexes $i$ and $j$ of the matrix element $D_{i,j}$ corresponds to $E_{\rm max}$ and $E_{\gamma}$, respectively.
The set of equations is given by
\begin{equation}
\begin{pmatrix}\sigma_{\rm{1}}\\\sigma_{\rm{2}}\\ \vdots \\ \sigma_N \end{pmatrix}_{\rm f}\\\mbox{}=
\begin{pmatrix}D_{ 11} & D_{ 12}& \cdots &\cdots &D_{ 1M} \\ D_{ 21} & D_{ 22}&
\cdots & \cdots &D_{ 2M} \\ \vdots &\vdots & \vdots & \vdots & \vdots \\ D_{ N1} & D_{ N2}& \cdots & \cdots &D_{ NM}\end{pmatrix}
\begin{pmatrix}\sigma_{1}\\\sigma_{2}\\ \vdots \\ \vdots \\\sigma_{M} \end{pmatrix}.
\label{eq:matrise_unfolding}
\end{equation}
Each row of $\bf{D}$ corresponds to a GEANT4 simulated $\gamma$
beam profile belonging to a specific measurement characterized by $E_{\rm max}$.  See Fig.~\ref{fig:GammaProfile} for a visual representation of the response matrix $\bf{D}$ for the case of $^{163}$Dy. It is clear that $\bf{D}$ is highly asymmetrical.
As mentioned, we have used $N=15$ beam energies when investigating $^{163}$Dy, but the beam profiles above $S_n$ is simulated for $M= 250$ $\gamma$ energies.
As the system of linear equations in Eq.~(\ref{eq:matrise_unfolding}) is under-determined, the true $\sigma$ vector cannot be extracted by matrix inversion. In order to find $\sigma$, we utilize a folding iteration method. The main features of this method are as follows:

\begin{itemize}

\item [1)] As a starting point, we choose for the 0th iteration, a constant trial function $\sigma^0$.
This initial vector
is multiplied with $\bf{D}$, and we get the 0th folded vector $\sigma^0_{\rm f}= {\bf D} \sigma^{0}$.
\item[2)] The next trial input function, $\sigma^1$, can be established by adding the difference of
the experimentally measured spectrum, $\sigma_{\rm{exp}}$, and the folded spectrum, $\sigma^0 _{\rm f}$,
to $\sigma^0$. In order to be able to add the folded and the input vector together, we first perform a spline
interpolation on the folded vector, then interpolate so that the two vectors have equal dimensions. Our new input vector is:

\begin{equation}
\sigma^1 = \sigma^0 + (\sigma_{\rm{exp}}-\sigma^0 _{\rm f}).
\end{equation} 

\item[3)] The steps 1) and 2) are iterated $i$ times giving
\begin{eqnarray}
\sigma^i_{\rm f} &=& {\bf D} \sigma^{i}
\\
\sigma^{i+1}     &=& \sigma^i + (\sigma_{\rm{exp}}-\sigma^i _{\rm f})
\end{eqnarray}
until convergence is achieved. This means that
$\sigma^{i+1}_{\rm f} \approx \sigma_{\rm exp}$ within the statistical errors.
In order to quantitatively check convergence, we calculate the reduced $\chi^2$ of $\sigma^{i+1}_{\rm f}$ and
$\sigma_{\rm{exp}}$ after each iteration.
Approximately four iterations are usually enough for convergence, which is defined when the reduced $\chi^2$ value approaches $\approx 1$.
\end{itemize}

It is important to stop the iteration
when the reduced $\chi^2$ starts to
be lower than unity. In principle, the iteration could continue until the reduced $\chi^2$ approaches zero,
but then large unrealistic fluctuations in $\sigma^i$ arises in order to reproduce the experimental
data points of $\sigma_{\rm exp}$ exactly, independently
of the individual error bars connected to each of these points.
To prevent the unfolding from introducing spurious fluctuations, we apply a smoothing factor of 200 keV, which corresponds to the average of the full-width half maximum (FWHM) of the $\gamma$ beams. 
In this way, we prevent the random fluctuations in the measured data from being amplified when using a spline in step 2) of the unfolding routine.

In order to give an estimate of the uncertainly in the unfolded cross sections, we have defined an upper limit of the monochromatic cross sections from Fig.\ref{fig:MonocrossBoth} by adding/subtracting the errors to the measured cross section values. This upper and lower limit is unfolded separately. 

\begin{figure}[]
\includegraphics[clip,width=1.\columnwidth]{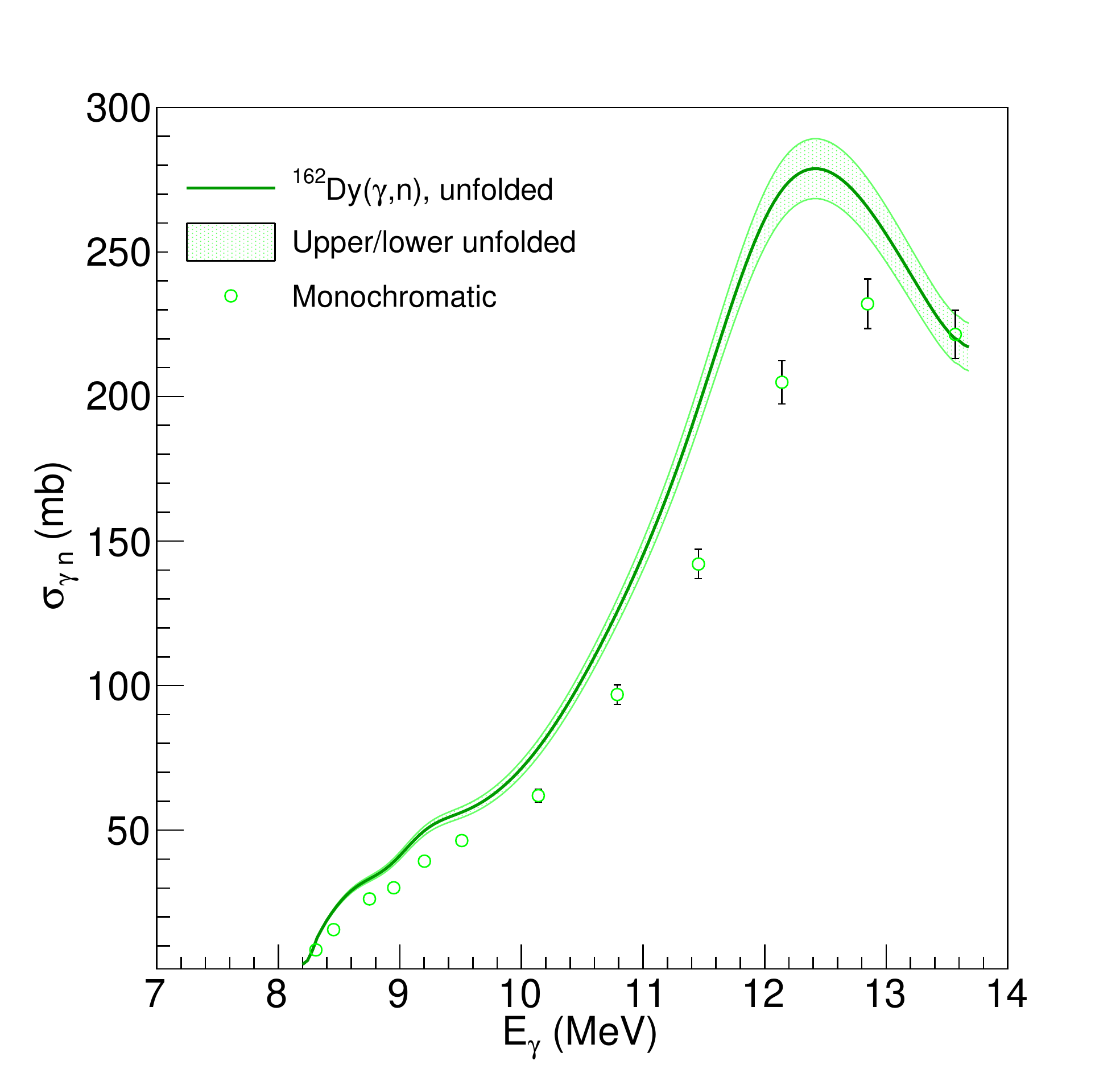}
\caption{(Color online) Cross sections of $^{162}\rm{Dy}$. The green, open circles are the monochromatic cross sections from Fig.\ref{fig:MonocrossBoth}. The green, shaded area display the unfolded cross section.}
\label{fig:Unfold_162}
\end{figure}

\begin{figure}[]
\includegraphics[clip,width=1.0\columnwidth]{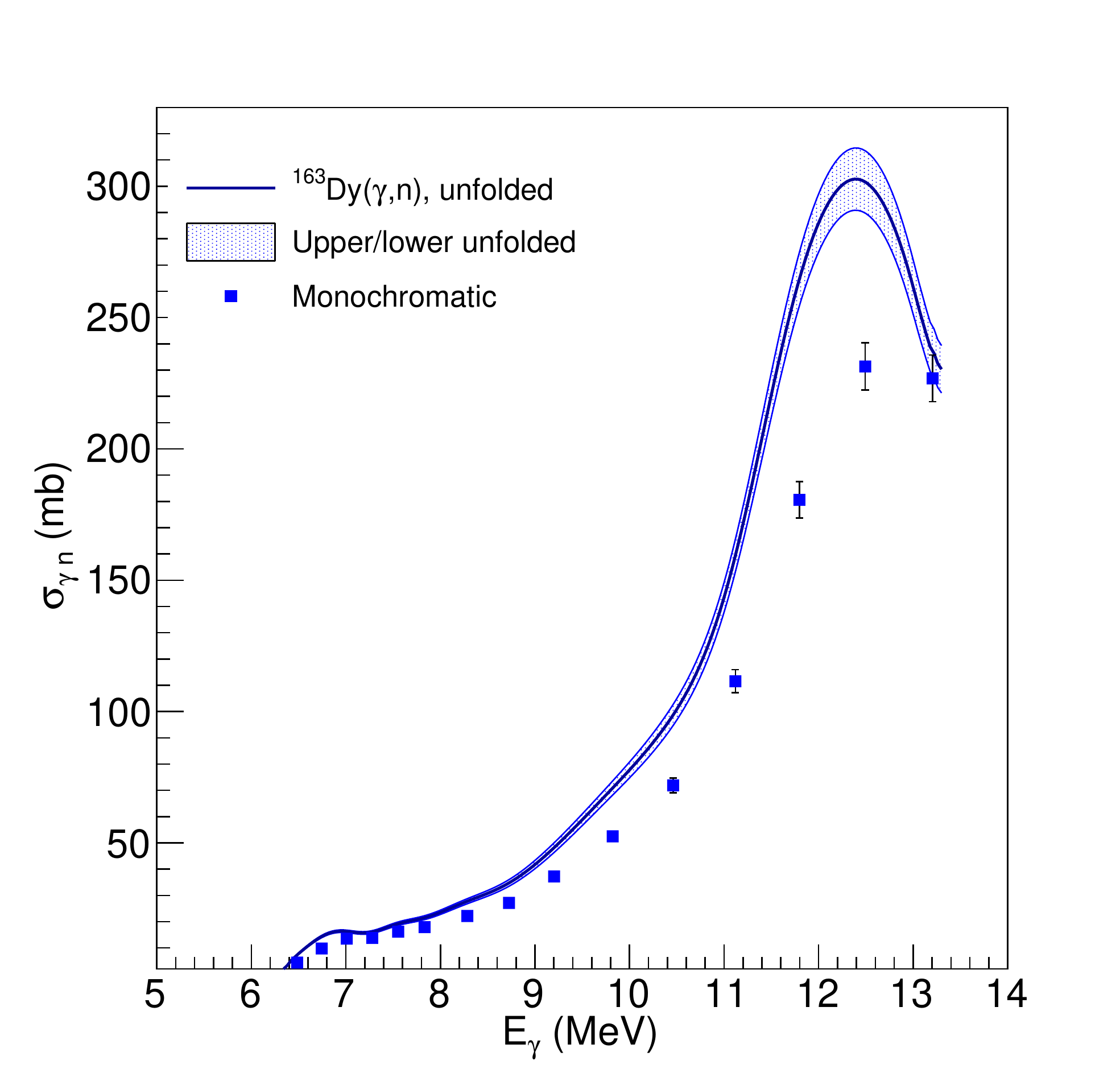}
\caption{(Color online) Cross sections of $^{163}\rm{Dy}$. The blue, filled squares are the monochromatic cross sections from Fig.\ref{fig:MonocrossBoth}. The blue, shaded area display the unfolded cross section.}
\label{fig:Unfold_163}
\end{figure}

Figures~\ref{fig:Unfold_162} and~\ref{fig:Unfold_163} show the resulting unfolded photo-neutron cross sections $\sigma(E_{\gamma})$ of $^{162,163}\rm{Dy}$.

\begin{figure}[]
\includegraphics[clip,width=1.0\columnwidth]{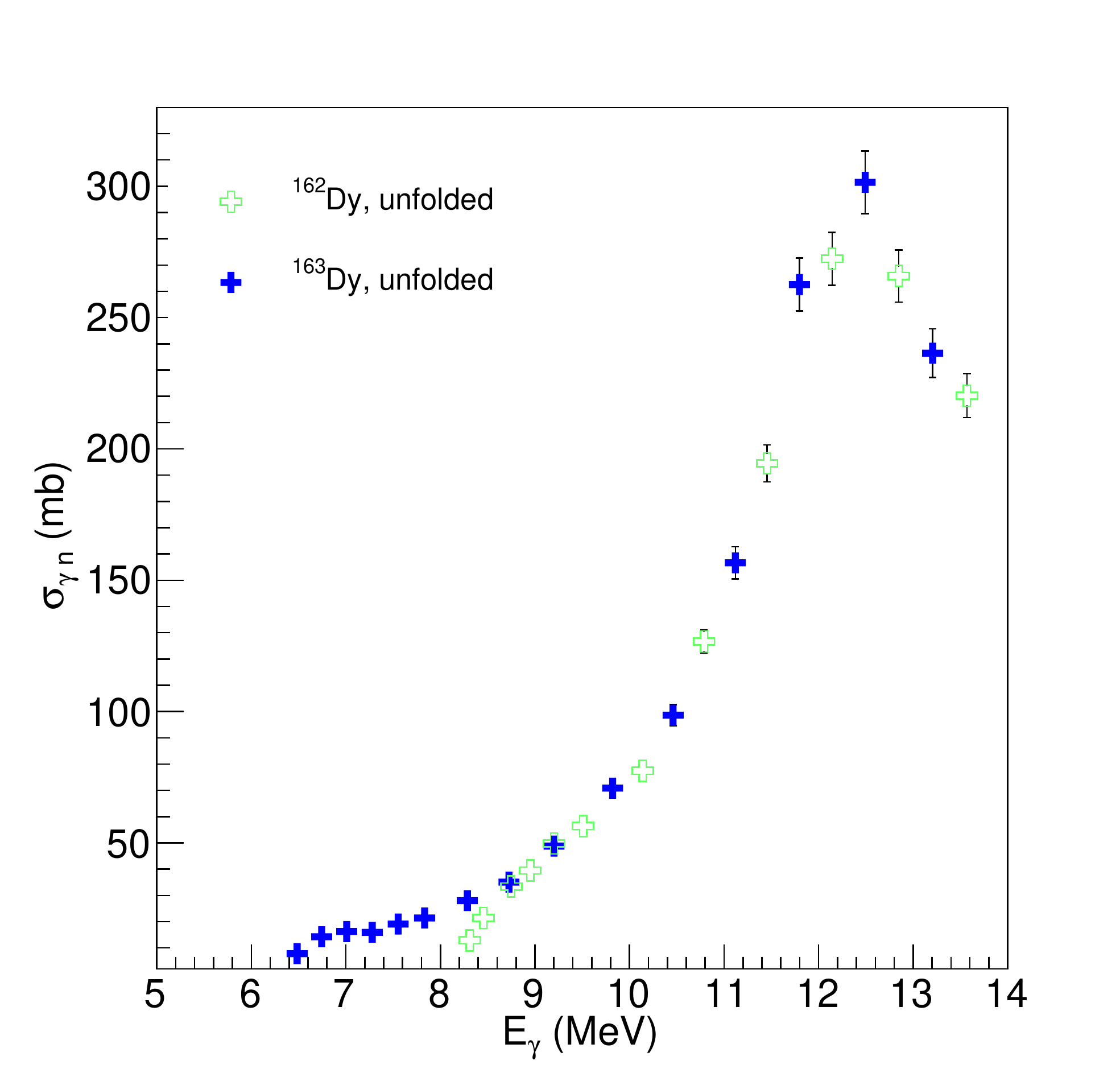}
\caption{(Color online) The recommended unfolded cross sections of $^{162,163}\rm{Dy}$. Here, the error bars represent the difference between the upper and lower limits shown in Figs.~\ref{fig:Unfold_162} and ~\ref{fig:Unfold_163}.}
\label{fig:Unfold_163162}
\end{figure}

In Fig.~\ref{fig:Unfold_163162} the two unfolded cross sections are evaluated at the maximum energies of the incoming $\gamma$ beams. The error bars represent the difference between the upper and lower limit of the unfolded cross sections.

\section{$^{160-164}$Dy revisited}

\label{sec_Oslo}
In the following, we reexamine previous data on $^{160-164}$Dy within the framework of the Oslo method.
The method is based on the analysis of particle-$\gamma$ coincidences obtained from transfer or inelastic reactions, where the
energy of the ejectile (and the reaction $Q$-value) uniquely determines the nuclear excitation energy, $E$.
These coincidence $\gamma$ spectra are organized as rows in a matrix of raw data, $R(E_{\gamma}, E)$.
The individual $\gamma$ spectra at each $E$
is then unfolded with the $\gamma$-detector response function~\cite{Gut96} giving the matrix $U(E_{\gamma}, E)$.
An iterative subtraction technique~\cite{Gut87} is performed on the $U$ matrix in order to obtain the
first-generation (or primary) $P(E_{\gamma}, E)$ matrix, containing the distribution of the first-emitted $\gamma$ rays for a given initial $E$. The next step is to
extract simultaneously the NLD, $\rho(E-E_{\gamma})$, and the $\gamma$SF, $f(E_{\gamma})$, by a least-$\chi ^2$ fit to the
$P$ matrix using the relation~\cite{Schiller00}
\begin{equation}
P(E_{\gamma}, E) \propto   \rho(E-E_{\gamma})E_{\gamma}^3 f(E_{\gamma}).
\end{equation}
This $\chi ^2$ minimization provides a unique solution, determining the functional form of $\rho(E-E_{\gamma})$ and $f(E_{\gamma})$.
The last step is to normalize the $\rho$ and $f$ functions to known external data. Further details of the Oslo
method and tests of various assumptions are given in Refs.~\cite{Schiller00,Lars11}.

\begin{table*}[]
\caption{Parameters used to extract level density and $\gamma$SF. Systematic uncertainties
are included in the errors for the recommended (R) normalization (see text). }
\begin{tabular}{c|ccr|cc|ccc|rr|c|c|c}
\hline
\hline
Nucleus&$S_n$& $a$           & $E_1$& $E_d$&$\sigma_d$&$\sigma_{\rm L}(S_n)$&$\sigma_{\rm R}(S_n)$&$\sigma_{\rm H}(S_n)$&$\delta$&$\alpha$&$D_0$  &   $\rho_{\rm R}(S_n)$&$\langle \Gamma_{\gamma}(S_n)\rangle$  \\
&(MeV)&(MeV$^{-1})$ & (MeV)&(MeV) &       &                      &             &             & (MeV)& &(eV)  &(10$^6$MeV$^{-1}$)       & (meV)   \\
\hline
$^{160}$Dy&8.576 & 18.78   &0.47  & 1.5  &  3.4(2)&            6.14        & 6.51        &   7.04      &$-0.56$ & 0.167 &1.84(41)$^{*}$  &    12.3(28)           &   112(20)$^{*}$\\
$^{161}$Dy&6.454 & 18.68   &$-0.55$ & 0.5  &  3.4(2)&            5.97        & 6.33        & 6.82        & $-0.71$ &$-0.059$ &27(5)           &    3.00(158)           &   108(10)      \\
$^{162}$Dy&8.197 & 18.50   &0.39  & 1.5  &  3.7(2)&            6.17        & 6.55        & 6.99        &$-0.54$ &0.153 &2.4(2)          &    6.67(110)           &   112(10)      \\
$^{163}$Dy&6.271 & 18.27   &$-0.53$ & 0.5  &  3.3(2)&            6.02        &  6.39       &  6.46       &0.15 &$-0.016$ &62(5)           &    1.33(29)           &   112(20)      \\
$^{164}$Dy&7.658 & 18.12   &0.31  & 1.5  &  3.6(2)&            6.18        & 6.55        & 6.84        &$-0.59$& 0.196 &6.8(6)          &    2.36(26)           &   113(13)      \\

\hline
\hline
\end{tabular}
\label{tab:parameters}

$^{*}$Taken from systematics.
\end{table*}

The present reanalysis is based on the raw NaI matrices obtained for $^{160-162}$Dy
in 2001 and 2003~\cite{Alex2001,Magne2003} and for $^{163-164}$Dy in 2010~\cite{Hilde2010,Hilde2012}
using the reactions ($^3$He,$^3$He$^\prime\gamma$) and ($^3$He,$\alpha\gamma$).
Since these data were first presented, new $\gamma$-ray response functions, improved data
software and normalization procedures have been implemented. In the present work, with the raw matrices ($R$) as basis, we
aim to perform a consistent treatment of the steps
needed to obtain the final NLDs and $\gamma$SFs for all $^{160-164}$Dy isotopes.

\subsection{Renormalization of NLDs}
\label{renorm:nld}
The nuclear spin distribution~\cite{Ericson1959}, which is usually expressed as
\begin{equation}
g(E,J) \simeq \frac{2J+1}{2\sigma_{J}^2(E)}\exp\left[-(J+1/2)^2/2\sigma_{J}^2(E)\right]
\label{eq:spindist}
\end{equation}
where $\sigma_{J}$ is the spin-cutoff parameter, plays an important role in obtaining $\rho(S_n)$ from known neutron-capture spacings, $D_0$.
The distribution also enters in the normalization of the $\gamma$SF to
reproduce the total average $\gamma$ width $\langle\Gamma_\gamma(S_n)\rangle$. This will be discussed briefly in the next section.

The Oslo group has used various empirical formulas for the estimation
of the $\sigma_{J}$ \cite{G&C1965,egidy2005,egidy2009}.
As was shown in e.g. Ref.~\cite{fransesca2015}, the various models may give large deviations.
The case is the same for the dysprosium isotopes, e.g. at $E=S_n$,
we find $\sigma_{J}$ values of $4.6 - 6.6$, and at $E= 0.5$ MeV the values are
$3.0 - 4.5$. Therefore, we adopt a more reliable approach as described in the following.

There is increasing evidence that the level density
follows closer the constant temperature
formula than the Fermi gas formula for excitations
above the pairing gap $2\Delta$ \cite{Moretto2014,CT2015}.
Assuming a constant temperature, $T$, the energy dependence in the expression $\sigma_{J}^2=\Theta T$ is
given by the moment of inertia, $\Theta$, which goes from
$\approx$ 50$\%$ of the rigid body moment of inertia ($\Theta_{\rm rigid}$) at the ground state
and approaches $\Theta_{\rm rigid}$ at $S_n$.
The moment of inertia is proportional to the number of quasiparticles
excited, which again is proportional to the excitation energy. Thus,
we assume $\sigma_{J}^2$ to be a linear function in $E$ by
\begin{equation}
\sigma_{J}^2(E)=\sigma_d^2 + \frac{E-E_d}{S_n-E_d}(\sigma_{J}^2(S_n)-\sigma_d^2).
\label{eq:sigE}
\end{equation}
The quantity $\sigma_d^2$ is determined from known discrete levels~\cite{nudat2017} at excitation
energy, $E_d$, where the level scheme is considered complete.

We observe that the level schemes of $^{160-164}$Dy are
close to complete in the excitation regions of 0.5 and 1.5 MeV for the
odd-mass and even-even dysprosiums, respectively. Thus, the $\sigma_d^2$ values can be reliably estimated.
The second data point at $S_n$ should approach a rigid moment of inertia,
which is assumed in Ref.~\cite{egidy2005}:
\begin{equation}
\sigma_{J}^2(S_n) = \Theta_{\rm rigid}\cdot T=0.0146 A^{5/3} \cdot \frac{1+\sqrt{1+4aU_n}}{2a},
\label{eqn:eb}
\end{equation}
where $A$ is the mass number, $a$ is the level density parameter, $U_n=S_n-E_1$ is the intrinsic excitation energy,
and $E_1$ is the energy shift parameter. To obtain an error band for $\sigma(S_n)$,
we introduce a reduction factor, $\eta$, in Eq.~(\ref{eqn:eb}) for $\Theta_{\rm rigid}$
with $\eta= 0.8$, and 0.9 for the low (L) and recommended (R) estimate, respectively.
The recommended value is consistent with theoretical estimates of $\sigma_{J} (S_n)$ in the dysprosium mass
region \cite{Rauscher1997,Capote2009,Uhrenholt2013,Alhassid2015,Grimes2016}.

\begin{figure}[tb]
\begin{center}
\includegraphics[clip,width=1.05\columnwidth]{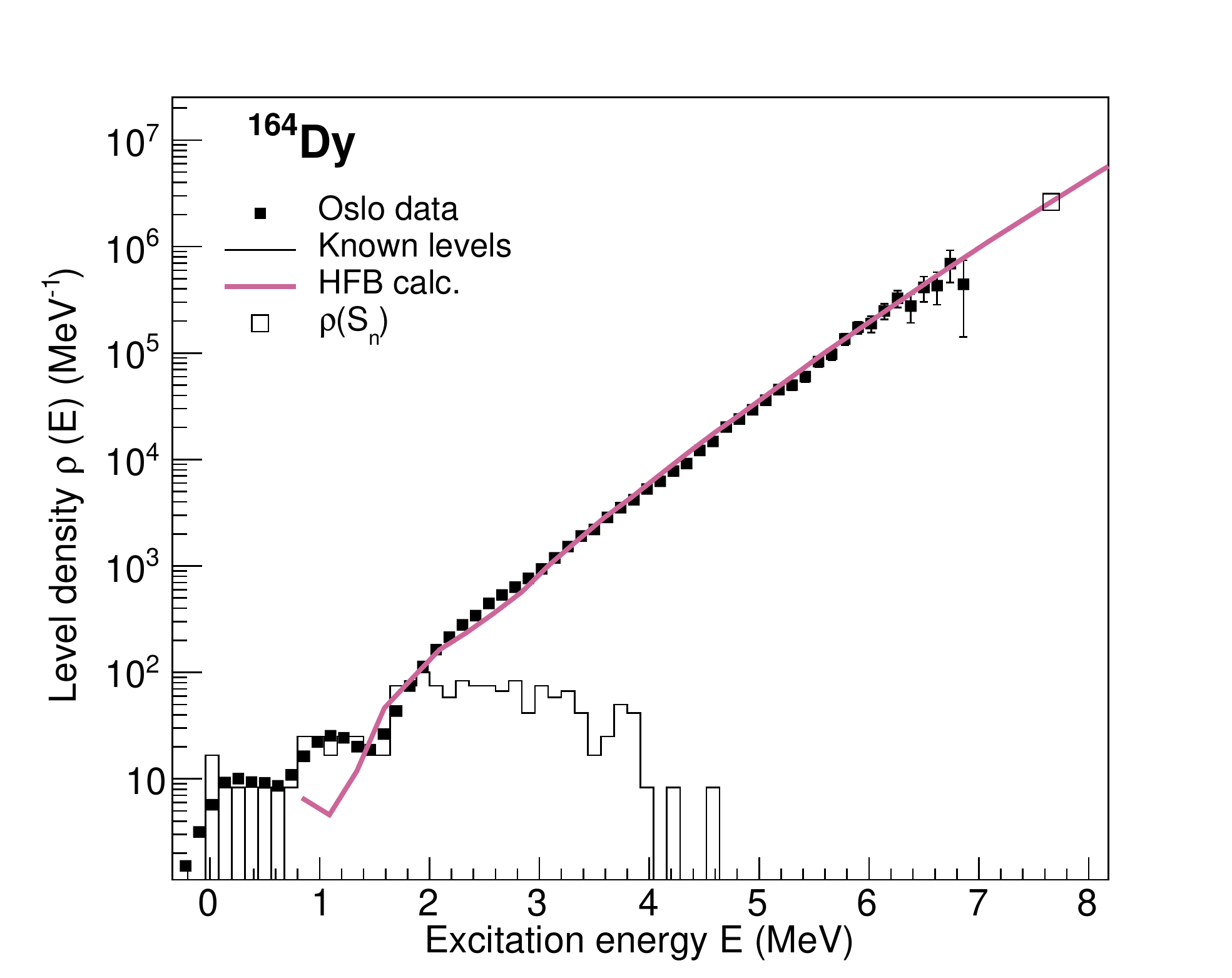}
\caption{(Color online) Normalization of the $^{164}$Dy NLD to the discrete levels and the HFB calculation from Ref.~\cite{Goriely2008}, which is adjusted to reproduce $D_0 = 6.8$ eV~\cite{RIPL3}.}
\label{fig:norm164DyHFB}
\end{center}
\end{figure}

\begin{figure}[tb]
\begin{center}
\includegraphics[clip,width=1.05\columnwidth]{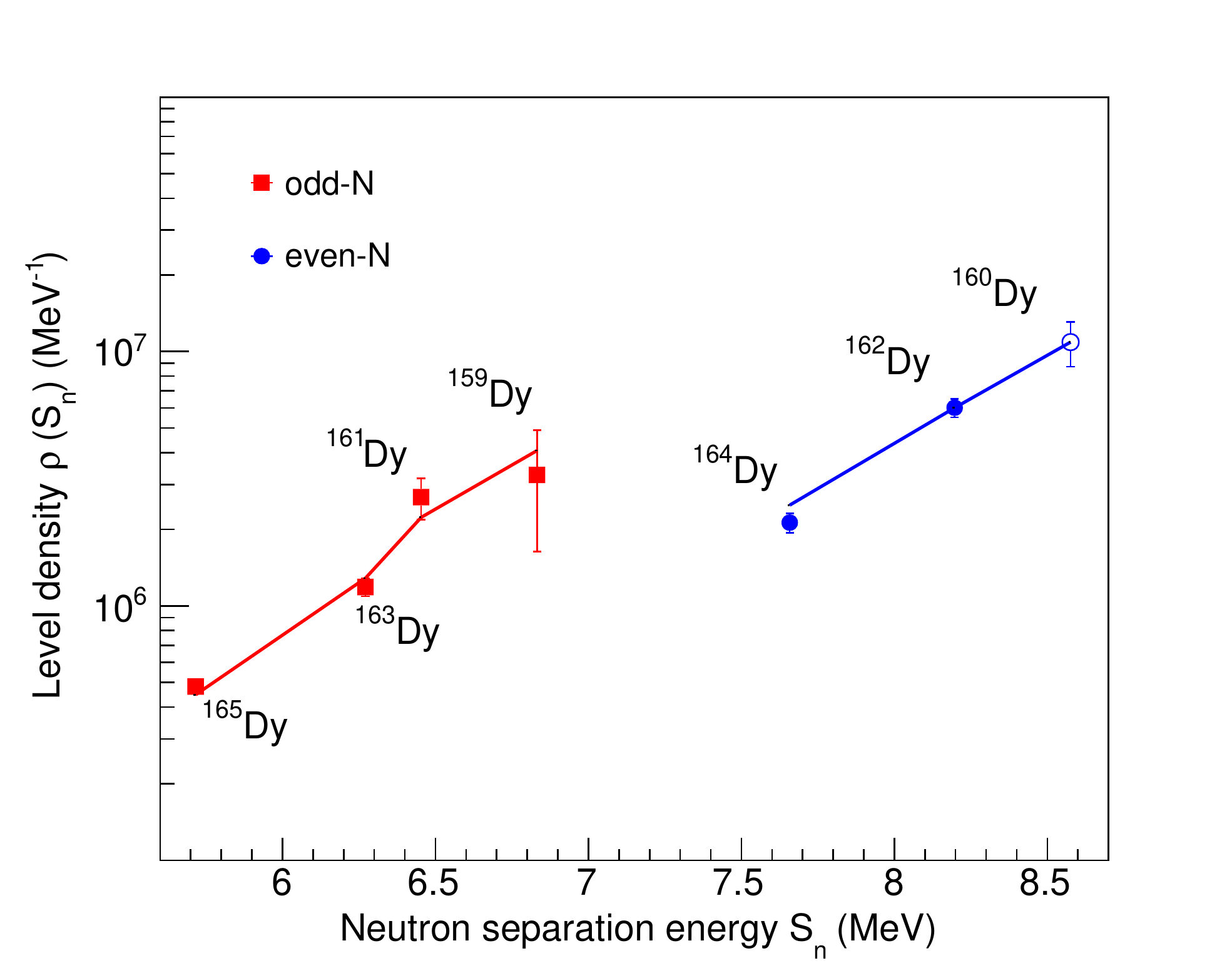}
\caption{(Color online) Level densities at the neutron separation energy
extracted from known $D_0$ values (filled symbols)~\cite{RIPL3}.
The systematics (solid lines) are taken from Ref.~\cite{egidy2005}
with all values scaled by a factor of 0.618. The estimate for $^{160}$Dy (open circle)
is taken from the scaled curve.}
\label{fig:syst}
\end{center}
\end{figure}

Further, for the upper limit (H), we apply the Hartree-Fock-Bogoliubov-plus-combinatorial (HFB) calculations of Ref.~\cite{Goriely2008}, adjusted to match the discrete levels and to reproduce the experimental $D_0$ values:
\begin{equation}
\rho_{\mathrm{HFB}}^{\mathrm{renorm}}(E) = \exp{[\alpha\sqrt{(E-\delta)}]} \rho_{\mathrm{HFB}}(E-\delta),
\end{equation}
where $\alpha$ is the slope correction and $\delta$ is an energy shift.
These calculations reproduce the overall shape of the experimental data very well for the even-even isotopes; the case of $^{164}$Dy is shown in Fig.~\ref{fig:norm164DyHFB}. 
For the odd isotopes, the shape of the NLD is less compatible with the data; we have chosen to still apply the spin distribution of Ref.~\cite{Goriely2008} to provide an anchor point at $S_n$ for the normalization of the NLD, but use the constant-temperature model for the interpolation as described in the following paragraph. 
The spin cutoff parameter $\sigma_\mathrm{H}(E)$ is extracted from the HFB calculations by a fit of Eq.~(\ref{eq:spindist}) for each excitation-energy bin.
Table~\ref{tab:parameters} lists the $D_0$ and  $\sigma_{J}$ values at $S_n$ used to determine
the level density $\rho(S_n)$ with $a$ and $E_1$ parameters taken from Ref.~\cite{egidy2005}.
The additional $E_d$ and $\sigma_d$ values are used to get the energy dependence according to Eq.~(\ref{eq:sigE}).
Also, the parameters $\alpha$ and $\delta$ are given.
The $D_0$ and $\langle \Gamma_{\gamma}(S_n)\rangle$
values are taken from $s$-wave neutron capture reactions reported in the  RIPL-3 compilation~\cite{RIPL3}.
As $^{159}$Dy is unstable, no neutron capture data is available for $^{160}$Dy and we therefore use arguments from systematics.
Figure \ref{fig:syst} demonstrates how $\rho(S_n)$ is estimated for $^{160}$Dy.
\begin{table}[tb]
\caption{Constant level density parameters extracted from fit to the experimental data of
the recommended normalization.}
\begin{tabular}{c|cc}
\hline
\hline
Nucleus&  $T_{\rm CT}$ & $E_0$  \\
&    (MeV)      &(MeV)   \\
\hline
$^{160}$Dy&0.61(2) & -1.01(21)  \\
$^{161}$Dy&0.59(2) & -1.97(31)  \\
$^{162}$Dy&0.61(1) & -1.02(16)  \\
$^{163}$Dy&0.59(4) & -1.67(58)  \\
$^{164}$Dy&0.60(1) & -0.78(15)  \\

\hline
\hline
\end{tabular}
\label{tab:ctparam}
\end{table}

\begin{figure*}[tb]
\begin{center}
\includegraphics[clip,width=2\columnwidth]{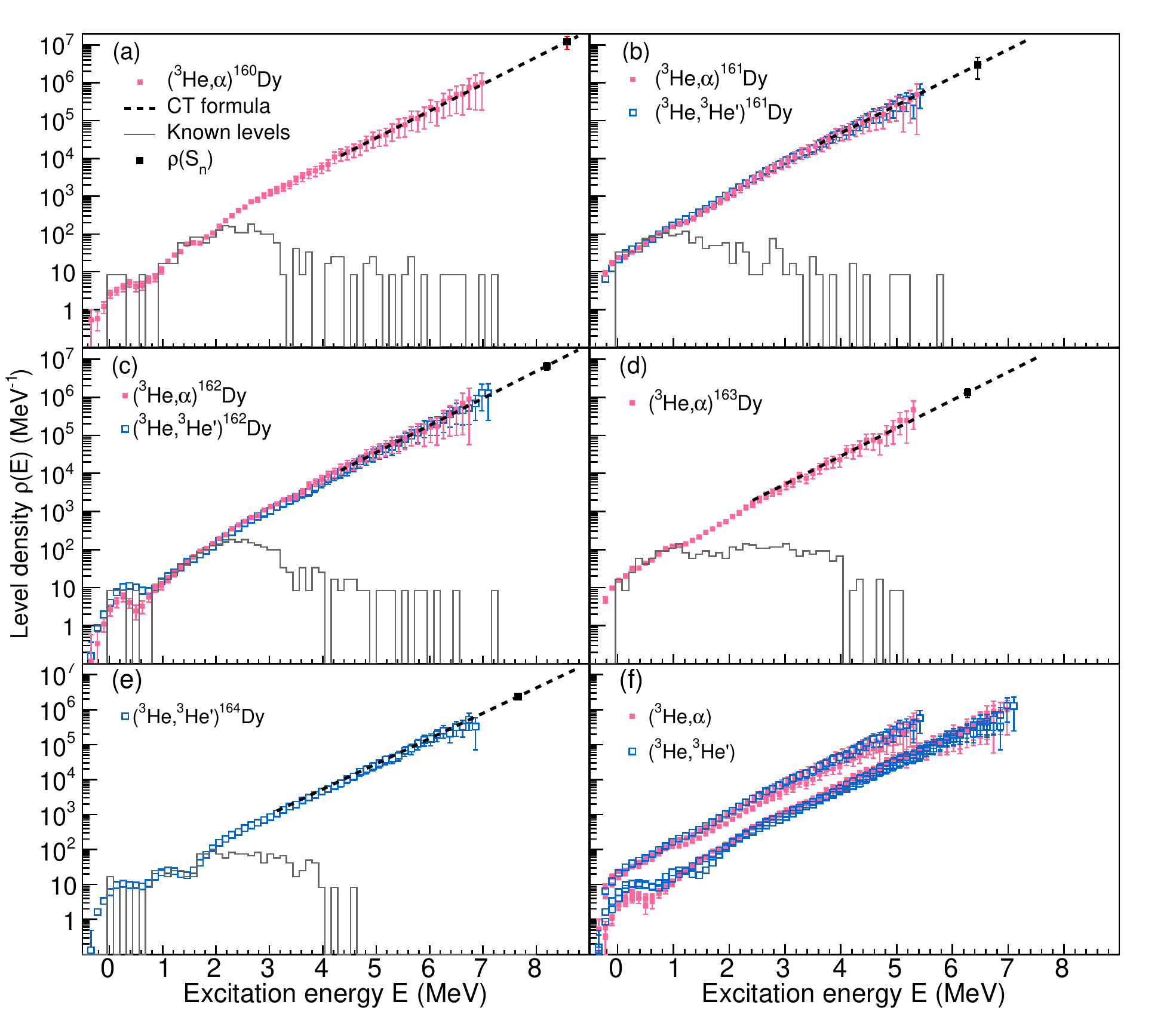}
\caption{(Color online) Level densities for $^{160-164}$Dy based on previous
data~\cite{Alex2001,Magne2003,Hilde2010,Hilde2012}. The pink (filled) and blue (open)
experimental data points are from the
($^3$He, $\alpha$) and ($^3$He, $^3$He$^\prime$) reactions, respectively.
The data are normalized
to the level density of known discrete levels~\cite{nudat2017} at low excitation energy $E$ (gray histogram) and
the level density $\rho(S_n)$ (black square) extracted from neutron resonance spacings $D_0$.
The dashed line shows the CT interpolation using Eq.~(\ref{eq:ct}) for the recommended normalization. Error bars include statistical and systematic errors (one standard deviation). }
\label{fig:rhotot}
\end{center}
\end{figure*}

The level densities obtained from the fit to the primary $\gamma$-ray matrix, $P$,
must be normalized to external data. For this purpose we use known discrete levels, the total
level density at $S_n$  based on the $D_0$ values, and the spin distribution $g(S_n,J)$
with spin cutoff parameters from Table~\ref{tab:parameters}. The interpolation of our
data points to the anchor point, $\rho(S_n)$, is obtained by the constant temperature (CT)
NLD formula~\cite{Ericson1959}
\begin{equation}
\rho_{\rm CT}(E)=\frac{1}{T_{\rm CT}}\exp\left({\frac{E-E_0}{T_{\rm CT}}}\right).
\label{eq:ct}
\end{equation}
Note that we choose to apply symmetric errors of $\rho(S_n)$, although the difference between the lower limit and the recommended normalization is typically 
smaller than for the upper limit. Hence, we use a conservative estimate for the error in $\rho(S_n)$, motivated by the rather large uncertainty in the spin-cutoff parameter. 

The final level densities are shown in Fig.~\ref{fig:rhotot}.
We note that there are no significant deviations
between the level densities obtained with the ($^3$He, $\alpha$) and ($^3$He, $^3$He$^\prime$) reactions.
The remarkable similarities between all seven nuclear level densities
in panel (f) reveal the same gross properties, which confirm
that these mid-shell dysprosium isotopes ($N=94-98$) have their
Fermi surfaces embedded between single particle orbits with
similar density and nuclear deformation.
We also observe that the nuclear level densities are close to a straight
line in the log-plot, in accordance with previous findings
using the Oslo method~\cite{Moretto2014,CT2015}.
The parallel level densities are also evident in Fig.~\ref{fig:rhotot} (f).
Here, we find that the odd-mass isotopes have $\approx 5$ times more levels compared to the even-even
isotopes. This clear difference in level density can be interpreted as the odd valence neutron bringing an additional entropy
of $\approx 1.7$ k$_{\rm B}$ into the system, rather independent on the number of paired nucleons~\cite{Magne2000}.

The nuclear temperature can be determined with
the constraints that $\rho_{\rm CT}(S_n)= \rho_{\rm R}(S_n)$.
We use a least-$\chi ^2$ fit of Eq.~(\ref{eq:ct}) to
the nuclear level density data points for $E > 2\Delta$,
where $\Delta \approx 12 A^{-1/2}$ is the pairing gap.
The fitted parameter values $T_{\rm CT}$ and $E_0$ with statistical uncertainties
are listed in Table~\ref{tab:ctparam}. All isotopes reveal the
same nuclear temperature within the statistical uncertainties.

We would like to stress that it makes no sense to fit neither the CT model nor the Fermi-gas model to data below $\approx 2\Delta$, where collective nuclear structure effects are predominant (rotation and vibration). 
Therefore, when discussing whether the level density behaves like a CT model or a Fermi gas, one must keep this in mind. When using data above $\approx 2\Delta$, the $\chi^2$ result of a fit to the Fermi gas model is significantly worse than a fit to the CT model, see Fig.9 in Ref.~\cite{CT2015}. 
The CT parameters given in Table~\ref{tab:ctparam} are extracted from a fit to our experimental data above $\approx 2\Delta$ for all the studied isotopes.

\subsection{Renormalization of $\gamma$SFs}
\label{renorm:gsf}
The $\gamma$SFs are normalized in such a way that the $\langle \Gamma_{\gamma}\rangle$
values of Table~\ref{tab:parameters} are reproduced by
\begin{eqnarray}
\langle\Gamma_\gamma(S_n)\rangle=&&\frac{D_0}{2\pi} \int_0^{S_n}{\mathrm{d}}E_{\gamma}2\pi E_{\gamma}^3 f(E_{\gamma})
\nonumber\\
&&\times \rho(S_n-E_{\gamma})\sum_{J_f}g(S_n - E_\gamma,J_f)
\label{eq:norm}
\end{eqnarray}
where the summation and integration run over all final levels with spin, $J_f$, that are accessible by $E1$ or $M1$
transitions with energy, $E_{\gamma}$.
The normalization procedures for the Oslo method are further described in Refs.~\cite{Schiller00,voinov2001}.

\begin{figure*}[tb]
\includegraphics[clip,width=2.1\columnwidth]{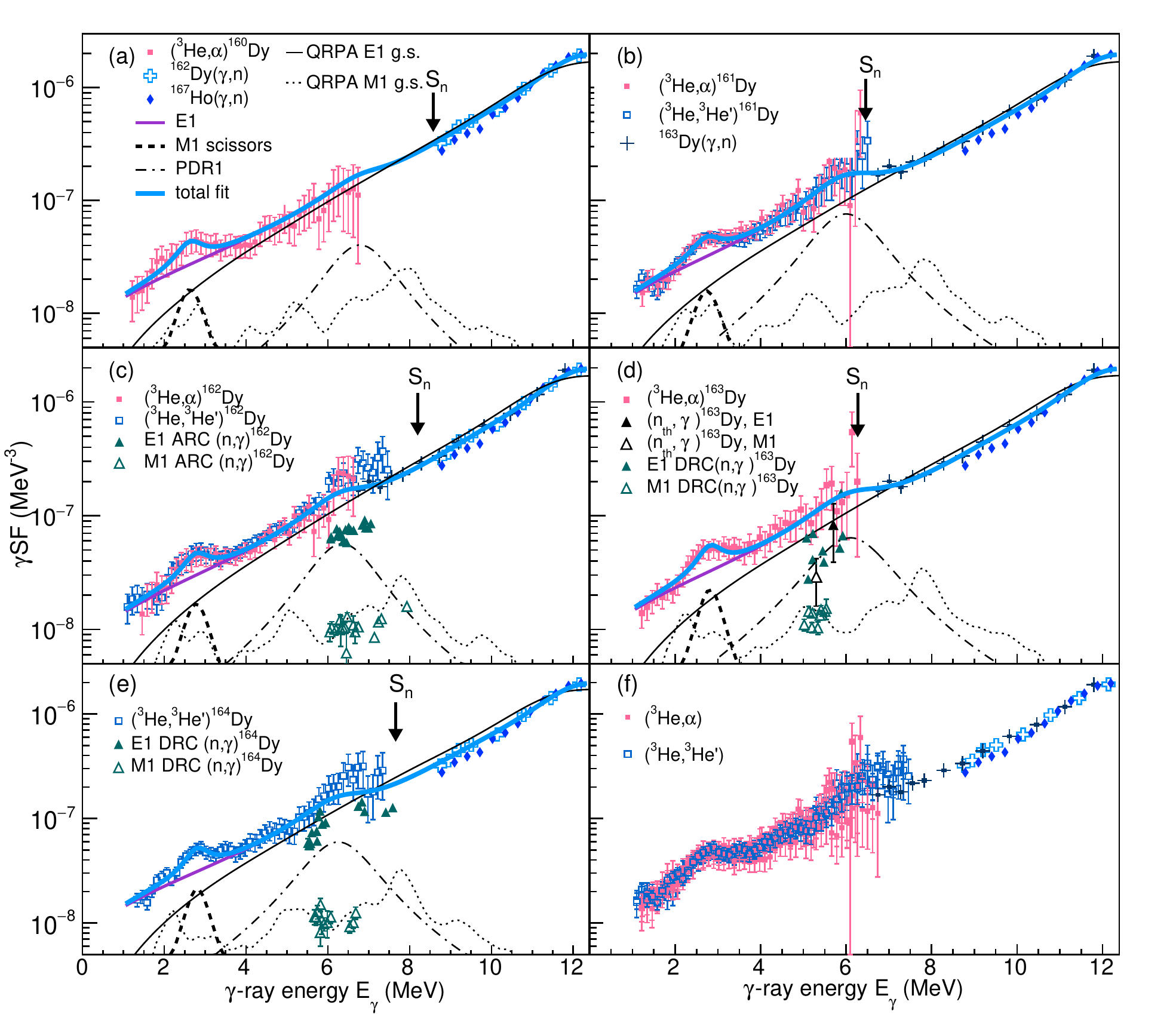}
\caption{(Color online). Comparison of Oslo $\gamma$SF data for  $^{160-164}\rm{Dy}$ with the present photo-neutron data for
$^{162,163}$Dy (blue, open and filled crosses, respectively) and the $^{165}$Ho($\gamma,n$) data (blue diamonds) from Berman \textit{et al.}~\cite{berman1969}.
The pink (filled) and blue (open) squares represent the reanalyzed Oslo data
from Refs.~\cite{Alex2001,Magne2003,Hilde2010,Hilde2012}. 
The solid, blue line is the total fit to the data as described in the text, with contributions from the GEDRs and PDRs (purple line)  
and the SR (dashed line). 
Calculations within the deformed-basis QRPA framework~\cite{martini2015,goriely2016} for ground-state excitations are shown and compared to data (solid, black line and dotted line for the $E1$ and $M1$ component, respectively). 
Re-evaluated $E1$ and $M1$ strengths obtained from average-resonance capture data~\cite{kopecky2017} are included in panels (c,d,e); in panel (e) we also
show $M1$ and $E1$ strength data from thermal ($n, \gamma$) reactions~\cite{RIPL3}.
Error bars include statistical and systematic errors (one standard deviation). Panel (f) shows all photoneutron data and Oslo data plotted together.}
\label{fig:gsf_all}
\end{figure*}

We would like to point out that an error in the normalization code was discovered in 2014: 
the spin weighting function, $g(S_n-E_\gamma,J_f)$, 
was applied with a wrong argument, instead of $S_n - E_\gamma$, only the $\gamma$-ray energy $E_\gamma$ was used. This resulted in an error in the absolute value of the $\gamma$SF of typically $\approx 30\%$. 
In addition, the spin cutoff parameters used previously and their dependence on excitation energy are different from the ones considered in the present analysis. 
Also, the back-shifted Fermi gas was applied for interpolating between $\rho(S_n)$ and our data points. 
All these factors lead to some differences between the previous normalizations and the present ones. 
The present normalizations are done in a consistent way with the same type of analysis for all isotopes, and we find that they are all very similar within the error bars (see Fig.~\ref{fig:gsf_all}f). 

 \section{Results}
 \label{sec_result}
 \subsection{Comparison of available strength-function data}
 \label{subsec:gsfdata}

According to 
the principle of detailed balance~\cite{blatt_weisskopf},
the upward and downward $\gamma$SF will correspond to each other.
The (upward) photo-neutron cross section $\sigma_{\gamma n}$ is connected to the (downward) $\gamma$SF by~\cite{RIPL3}
\begin{equation}
f(E_{\gamma})=\frac{1}{3\pi^2\hbar^2c^2}\frac{\sigma_{\gamma n}(E_{\gamma})}{E_{\gamma}},
\label{eq:crossgsf}
\end{equation}
where the constant $1/3\pi^2\hbar^2c^2 = 8.674 \cdot 10^{-8}$ mb$^{-1}$ MeV$^{-2}$.  Using this relation, one can compare the $\gamma$SFs from the newly-measured ($\gamma$, n) data with the reanalyzed
Oslo data. 
Note that Eq.~\ref{eq:crossgsf} holds only when the neutron channel in the photoneutron data dominates. 
In the vicinity of $S_n$, the competing $\gamma$ emission needs to be taken into account through the Hauser-Feshbach formalism.

Figure~\ref{fig:gsf_all} shows the $^{162,163}$Dy($\gamma$, n) data for
energies above $S_n$, while the reanalyzed Oslo data covers energy regions below $S_n$. Previous measurements on the $^{165}$Ho($\gamma,n)$ cross section~\cite{berman1969} are also shown. 
In the cases of $^{162,163}$Dy, the two distinct types
of data and normalization procedures match remarkably well.
Moreover, re-evaluated $E1$ and $M1$ strengths obtained from average-resonance capture data~\cite{kopecky2017} are plotted in 
Fig.~\ref{fig:gsf_all}c-e.
In Fig.~\ref{fig:gsf_all}f, all the experimental photo-neutron and Oslo data are plotted together, showing a consistent behavior. 

\subsection{Comparison with models}
\label{subsec:models}
The solid blue curve of Fig.~\ref{fig:gsf_all} shows a fit to
the $\gamma$SF data by the sum of five
functions:
the double-humped GEDR, a pygmy dipole resonance (PDR1) at a lower centroid of $\approx 6-7$ MeV,
a second PDR (PDR2) at a rather high energy centroid of $\approx 11$ MeV, and the scissors resonance (SR):
\begin{equation}
f=f_{\rm E1,1} + f_{\rm E1,2}  + f_{\rm PDR1} + f_{\rm PDR2} + f_{\rm SR}.
\label{eq:sumGSF}
\end{equation}

The PDRs were included to get a good reproduction of the photoneutron data in the
$E_{\gamma} = 8-11$ MeV region, and also to describe the Oslo data reasonably well in the $E_\gamma \approx 6-7$-MeV  region. Although the Oslo $\gamma$SFs have rather large error bars for high $E_\gamma$, the overall trend indicates the presence of a PDR centered between $E_\gamma \approx 6-7$ MeV. 
The electromagnetic character is likely electric, considering the data from Ref.~\cite{kopecky2017}. 
Also, NRF data from Volz \textit{et al.}~\cite{Volz2006} on $N=82$ isotones, amongst them $^{144}$Sm, show a concentration of $E1$ strength between $E_x \approx 5.5-6.5$ MeV attributed to a PDR.  
Based on the weak $M1$ strengths found in Ref.~\cite{kopecky2017} between $E_\gamma \approx 5-7$ MeV, which are about a factor of $\approx 6-8$ lower than the $E1$ strength, we did not add an $M1$ spin-flip resonance to the fit. 

For the GEDR we have used the GLO model~\cite{ko90,RIPL3} with two components:
\begin{align}
&f_{\rm E1,i}(E_{\gamma}) = \frac{1}{3\pi^2\hbar^2c^2}\sigma_{\rm E1,i}\Gamma_{\rm E1,i}   \nonumber \\
& \times \left[ \frac{ E_{\gamma} \Gamma(E_{\gamma},T_f)}{(E_\gamma^2-\omega_{\rm E1,i}^2)^2 + E_{\gamma}^2 \Gamma^2(E_{\gamma},T_f)}
+ 0.7 \frac{\Gamma(E_{\gamma}=0,T_f)}{\omega_{\rm E1,i}^3}\right]
\label{eq:GLO}
\end{align}
with a $\gamma$ width of
\begin{equation}
\Gamma(E_{\gamma},T_f) = \frac{\Gamma_{E1,i}}{E_{E1,i}^2} (E_{\gamma}^2 + 4\pi^2 T_f^2).
\end{equation}
The index $i$ gives the first and second part of the resonance.
The PDR and SR structures have been described by the SLO model~\cite{RIPL3}:
\begin{equation}
f_{\rm SLO}(E_{\gamma}) = \frac{1}{3\pi^2\hbar^2c^2}\frac{\sigma_{\rm SLO} E_{\gamma} \Gamma_{\rm SLO}^2}{(E_\gamma^2-\omega_{\rm SLO}^2)^2 + E_{\gamma}^2 \Gamma_{\rm SLO}^2},
\label{eq:SLO}
\end{equation}
with resonance parameters $(\omega_{\rm SLO}, \sigma_{\rm SLO}, \Gamma_{\rm SLO})$ given by the individual PDR and SR structures.

The present $^{162,163}$Dy($\gamma$,n) data provide information on the resonance parameters up to $\approx 13.5$ MeV, including the first peak of the double-humped GEDR. 
For the second peak around 15.2 MeV, we make use of the $^{165}$Ho($\gamma,n$) data by Berman \textit{et al.}~\cite{berman1969}. 
Thus, for the upper $E1$ part, $f_{E1} = f_{\rm E1,1} + f_{\rm E1,2}  +  f_{\rm PDR2}$, we have fitted simultaneously the $^{162,163}$Dy($\gamma$,n) and  $^{165}$Ho($\gamma,n$) data (from $E_\gamma \approx 13-16.5$ MeV) in the region $E_\gamma = 8.0-16.5$ MeV. 
Further, we make use of the constant temperature determined by the fit of the level densities, so that $T_f = T_{CT}$ from Table~\ref{tab:ctparam}. 
With this strategy we estimate the $E1$ component of the $\gamma$SF for the Oslo data. 
The resonance parameters for the GEDR1, GEDR2 and PDR2 resonances
are listed in Table~\ref{tab:GDR_param}. With the temperature parameter determined from the level density, nine free parameters are included in the fit. Note that we have also tested the case where $T_f$ is a free (but constant-value) parameter as well; this gives slightly lower temperatures of $T_f \approx 0.5$ MeV. 

Keeping the $E1$ parameters fixed except for PDR1, we  fit the Oslo $\gamma$SF data in the range $E_\gamma \approx 1.5 - 8.9$ MeV to the function in Eq.~(\ref{eq:sumGSF}) to determine the PDR1 and SR parameters (six free parameters). 
In the cases where there are data from two different reactions, both data sets are included simultaneously in the fit. 
We have not applied any constraints on the fit parameters, except for $^{160}$Dy where we have used a fixed width of the PDR1 resonance of $\Gamma_{\mathrm{PDR1}} = 1.9$ MeV (taken from the fit of the ($^3$He,$\alpha$)$^{161}$Dy and ($^3$He,$^3$He$^\prime$)$^{161}$Dy data). 
The reason for locking this parameter is the large fluctuations in the data for $E_\gamma > 5$ MeV for this nucleus. The resulting PDR1 parameters are given together with the other $E1$ contributions in Table~\ref {tab:GDR_param}. 

The SR parameters are given in Table~\ref{tab:SR}. 
We find that the centroids, $\omega_{\rm SR}$, are very well determined in all cases. 
Also the width, $\Gamma_{\rm SR}$, and peak cross section, $\sigma_{\rm SR}$, are reliably estimated, although with larger error bars.

\begin{table*}[bt]
\caption{Resonance parameters used for the $E1$ $\gamma$SF.}
\begin{tabular}{cccccccccccccc}
\hline
\hline
Nucleus&$\omega_{E1,1}$&$\sigma_{E1,1}$&$\Gamma_{E1,1}$&$\omega_{E1,2}$&$\sigma_{E1,2}$&$\Gamma_{E1,2}$&$T_f$&$\omega_{\rm PDR1}$&$\sigma_{\rm PDR1}$&$\Gamma_{\rm PDR1}$      &$\omega_{\rm PDR2}$&$\sigma_{\rm PDR2}$&$\Gamma_{\rm PDR2}$ \\
&(MeV)  &      (mb)     &      (MeV)    &   (MeV)  &     (mb)      &    (MeV)      &(MeV)&    (MeV)    &      (mb)        &          (MeV)                               &    (MeV)    &      (mb)        &          (MeV)       \\ \hline
$^{160}$Dy& 12.67(5)    &      264(16)      &       3.0(2)     &     15.20(3) &      176(8)      &     2.2(2)       & 0.61 &               6.9(1)   &      3.2(2)        &       1.9$^{*}$&            10.6(8)   &      30(12)         &        4.9(10)           \\
$^{161}$Dy& 12.68(5)    &      262(17)      &       3.0(2)     &     15.20(3) &      175(8)      &     2.2(2)       & 0.59 &              6.08(9)   &      5.3(10)        &        1.9(3)&         10.6(8)   &      30(12)         &        5.0(10)              \\
$^{162}$Dy& 12.67(5)    &      264(16)      &       3.0(2)     &     15.20(3) &      176(8)      &     2.2(2)       & 0.61 &              6.42(9)   &      4.2(5)        &        1.9(2) &           10.6(8)   &      30(12)         &        4.9(10)            \\
$^{163}$Dy& 12.68(5)    &      262(18)      &       3.0(2)     &     15.20(3) &      175(8)      &     2.2(2)       & 0.59 &             6.19(19)   &      4.5(9)        &        2.1(3) &          10.6(8)   &      30(12)         &        5.0(10)             \\
$^{164}$Dy& 12.68(5)    &      263(16)      &       3.0(2)     &     15.20(3) &      175(8)      &     2.2(2)       & 0.60 &             6.33(14)   &      4.3(7)        &        1.9(3) &          10.6(8)   &      30(11)         &        5.0(10)             \\

\hline
\hline
\end{tabular}
\label{tab:GDR_param}

$^{*}$Taken from $^{161}$Dy.
\end{table*}

\begin{table*}[tb]
\caption{Scissors resonance fit parameters and $B_{\rm SR}$ strengths.}
\begin{tabular}{r|c|lllcl}
\hline
\hline

Nucleus&Deformation&\multicolumn{5}{c}{Experiment}\\
\hline

&$\delta$& $\omega_{\rm SR}$& $\sigma_{\rm SR}$ & $\Gamma_{\rm SR}$&$B_{\rm SR}^{(a)}$&$B_{\rm SR}^{(b)}$ \\
&        &      (MeV)       &    (mb)           &     (MeV)        & ($\mu_N^2$)& ($\mu_N^2$) \\
\hline
$^{160}\rm{Dy}$ &0.320&2.66(12)&0.50(17)& 0.79(33)    &4.8(26) & 1.7(10) \\
$^{161}\rm{Dy}$ &0.323&2.78(7) &0.50(7) & 0.79(18)    &4.6(12) & 2.1(6)  \\
$^{162}\rm{Dy}$ &0.325&2.81(8) &0.54(12) & 0.76(21)   &4.8(17) & 2.3(8)  \\
$^{163}\rm{Dy}$ &0.327&2.84(15)&0.73(19)& 0.69(25)    &5.8(26) & 3.1(14) \\
$^{164}\rm{Dy}$ &0.329&2.83(8) &0.69(14)& 0.69(18)    &5.5(18) & 2.8(9)  \\
\hline
\hline
\end{tabular}
\\
\label{tab:SR}
$^{(a)}$ Limits of integration $0-10$ MeV. $^{(b)}$ Limits of integration $2.7-3.7$ MeV.
\end{table*}

In Fig.~\ref{fig:gsf_all}, we compare the data with recent calculations within the quasiparticle random-phase approximation (QRPA) using an axially-symmetric deformed basis~\cite{martini2015,goriely2016}. 
The QRPA calculations probe the $E1$ and $M1$ strength built on the ground state. 
It is very interesting to see that the $E1$ QRPA calculations are in excellent agreement with the fitted $E1$ component down to $E_\gamma \approx 4.5$ MeV, indicating that the $E1$ strength built on the ground state is a good proxy for the $E1$ strength in the quasi-continuum as well. 
Recent shell-model calculations by Sieja~\cite{sieja2017} demonstrate that the low-energy $E1$ part probably attains a flat, constant strength towards $E_\gamma \rightarrow 0$ MeV. 
This brings further support to the $E1$ component extracted from the fit to the ($\gamma,n$) data.  

The deformed-basis $M1$ QRPA calculations clearly display significant structures. 
A splitting of the SR-like strength between $E_\gamma = 1.5-3.5$ MeV is particularly intriguing. 
Such a splitting has previously been experimentally observed in the actinide region~\cite{guttormsen2014}; however it is not clear from the present Dy data whether this is also the case here. 

\subsection{The $M1$ scissors resonance}

\begin{figure}[tb]
\includegraphics[clip,width=1.\columnwidth]{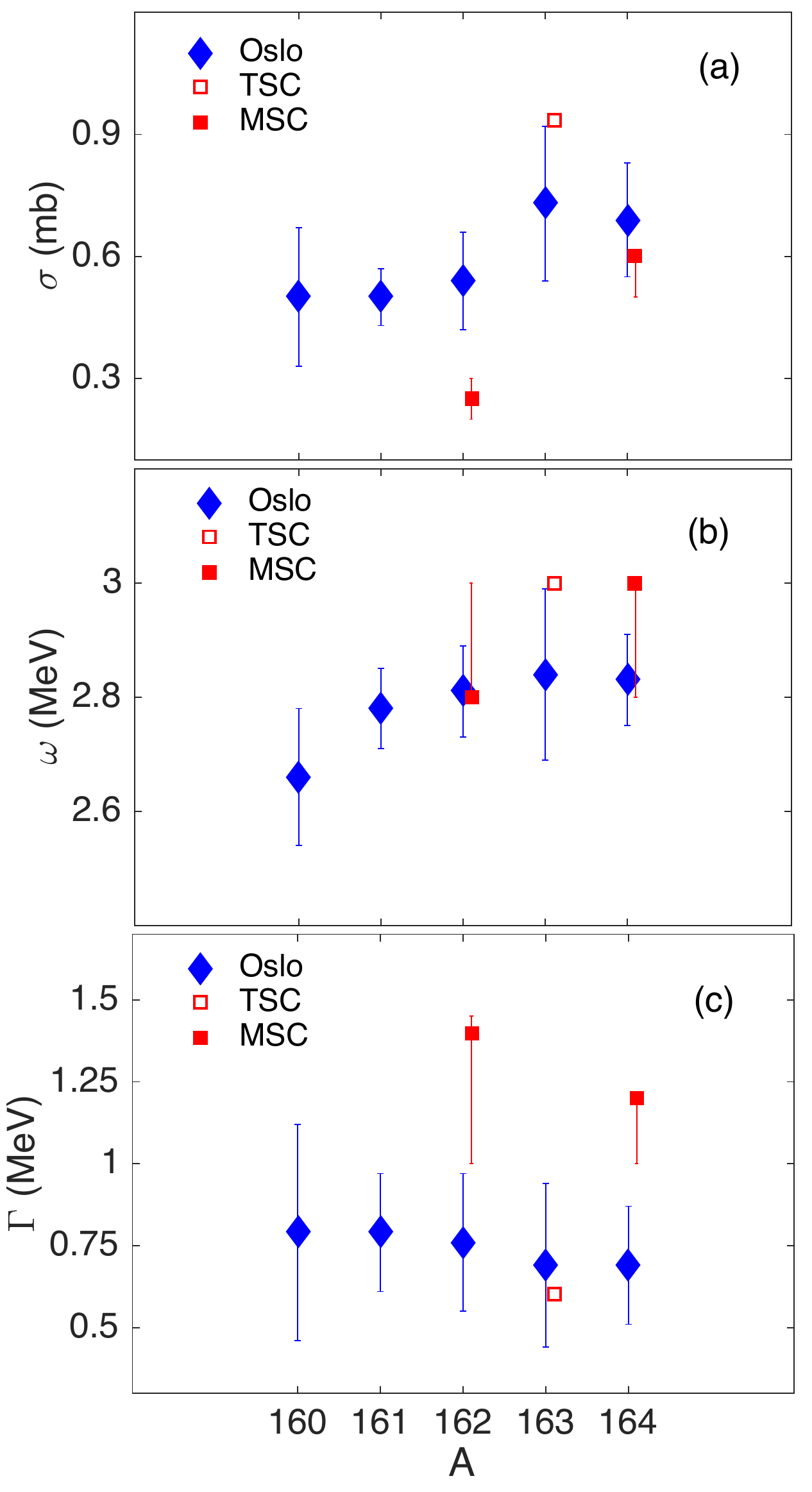}
\caption{(Color online). Comparison of the SR resonance parameters: (a) peak cross sections, (b) centroids and (c) widths from this work (blue diamonds) with reported parameters from Ref.~\cite{Krticka_letter} (red open squares) and Ref.~\cite{valenta2017} (red filled squares).}
\label{fig:ScissorParameters}
\end{figure}

We now turn to the SR and would like to compare our present results with other experimental findings and the QRPA calculations. 
The systematics of the SR parameters are shown in Fig.~\ref{fig:ScissorParameters}, where the Oslo parameters are taken from Table~\ref{tab:SR}. 
The present results are very similar within the uncertainties, 
which is expected because the deformation of these isotopes 
is about the same. However, there is 
a tendency towards higher $\omega_{\rm SR}$
for the heavier isotopes. As the 
centroid is directly proportional to the deformation parameter $\delta$, this 
might indicate a slightly larger
deformation for $^{163,164}$Dy.
Also, our results for the SR parameters of $^{163}$Dy compare well with parameters published in Ref.~\cite{Krticka_letter}, 
although it is difficult to assess the degree of agreement, 
because the authors of Ref.~\cite{Krticka_letter} did not report any errors.

We also compare with parameters deduced from the MSC analysis in Ref.~\cite{valenta2017}. 
The peak cross section of $^{162}$Dy reported in~\cite{valenta2017} is considerably smaller than the peak cross extracted from the Oslo type experiment, but for $^{164}$Dy the results agree within the error bars. The resonance widths reported in~\cite{valenta2017} are almost a factor 2 larger then our reported widths. It is interesting to note the difference in SR widths deduced from the recent MSC data on $^{162,164}$Dy and the TSC data on $^{163}$Dy.         

To determine the experimental summed SR strength from our data, we numerically integrate Eq.~(\ref{eq:SLO}) by\footnote{Note that in previous works, the expression $B_{\rm SR}=\frac{9\hbar c}{32\pi^2}\left( \frac{\sigma_{\rm SR}\Gamma_{\rm SR}}{\omega_{\rm SR}}\right)$ has been used to estimate the integrated SR strength. 
This formula gives $\approx 10\%$ higher strength than integrating the SR function as given in Eq.~(\ref{eq:integrated}).} 
\begin{equation}
B_{\rm SR} = \int \frac{dB_{\rm SR}}{dE_\gamma} dE_\gamma = \frac{27(\hbar c)^3}{16\pi} \int f_{{\rm SR}}(E_\gamma) dE_\gamma,
\label{eq:integrated}
\end{equation}
where  $27(\hbar c)^3/16\pi = 2.5980 \cdot 10^8$ $\mu_N^2$MeV$^2$. 
When comparing to existing NRF data, we observe varying summing conventions. 
In Ref.~\cite{Pietralla1995}, the excitation energy summing interval for Z$<$68 nuclei is set to $2.7 - 3.7$ MeV and for higher $Z$, the range is typically $2.4-3.7$ MeV.
In addition, transitions that are identified as $M1$ spin-flip type from inelastic proton scattering~\cite{Frekers1989} are omitted from the sum.  

\begin{figure}[tb]
\includegraphics[clip,width=1.\columnwidth]{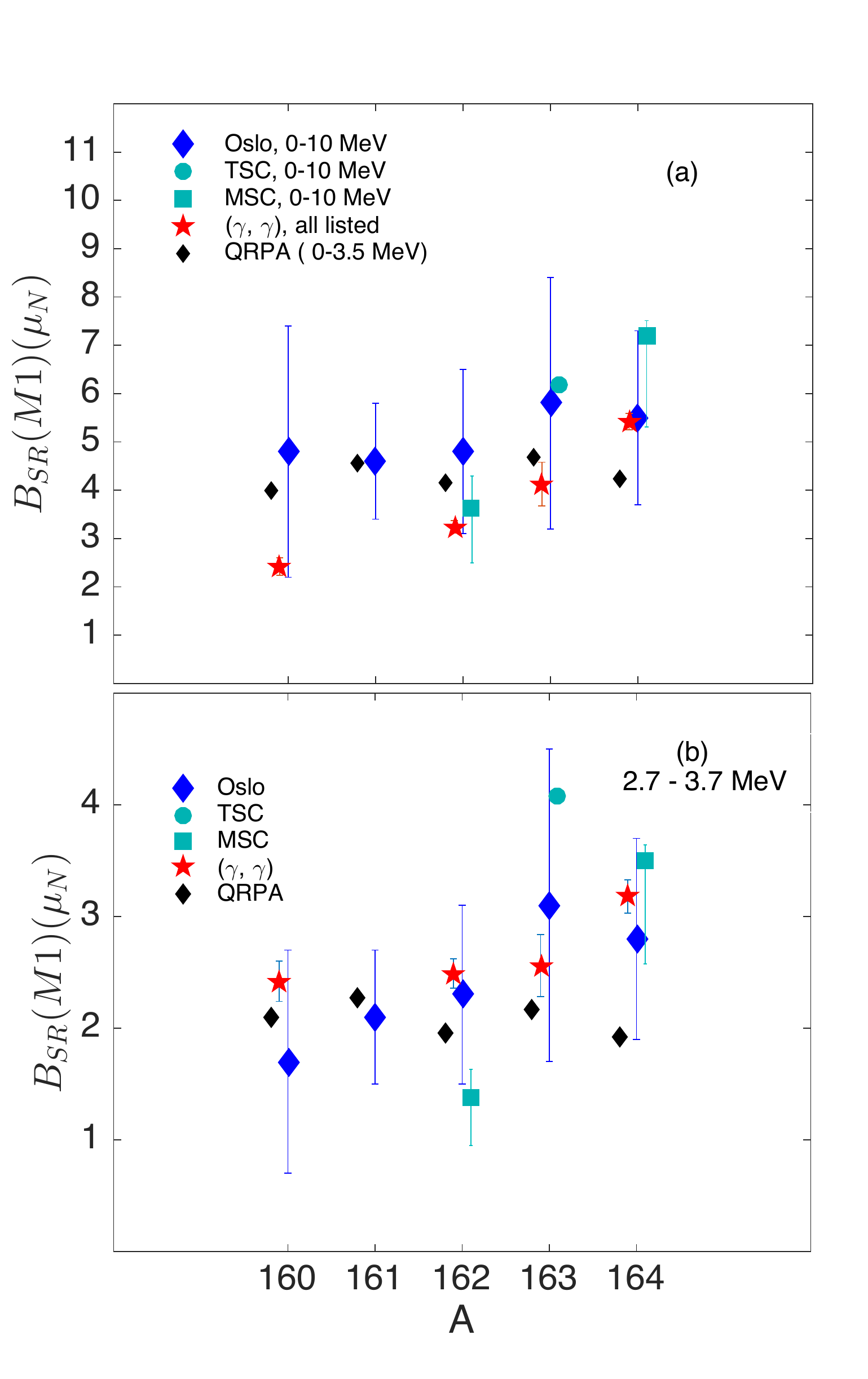}
\caption{(Color online) Comparison of the integrated SR strength from the present data sets (blue diamonds), the TSC measurement~\cite{Krticka_letter} (green filled squares) and the MSC results (green filled circles). The error bars of the MSC data are taken from Ref.~\cite{valenta2017} and the recommended values (represented by green filled circles) are from ~\cite{valentaPrivateCommunication}. The NRF measurements for $^{160}$Dy are from Ref.~\cite{Wesselborg}, for $^{162,164}$Dy from~\cite{MargrafDy} and for $^{163}$Dy from~\cite{Nord}. All NRF measurements are represented by red stars. The QRPA calculations (black diamonds) are from Ref.~\cite{goriely2016}. In (a) we show the total integrated/summed SR strength for all energies in the relevant SR energy range and in (b) we show the integrated/summed SR strength for the energy interval $2.7 - 3.7$ MeV. The SR strengths for $^{160,162,164}$Dy are the evaluated values of the NRF measurements from Ref.~\cite{Pietralla1995}. Note the different scales on the axes.}
\label{fig:BM1_all}
\end{figure}

If we integrate Eq.~(\ref{eq:integrated}) over all transition energies, we find a total, summed SR strength of $4.6(12)-5.8(26)$ $\mu_N^2$. 
We observe that treating $T_f$ as a free parameter in the fit will lead to larger summed SR strengths.  
The present fit strategy gives about 40\% higher summed SR strengths  than the reported NRF results. 
However, if we apply the NFR energy limits to Eq.~(\ref{eq:integrated}), we obtain excellent agreement with the NRF results as shown in Fig.~\ref{fig:BM1_all}b. 
It is interesting to note that $\approx 40 - 60 \%$ of our measured SR strength lies in the energy region below 2.7 MeV. In traditional NRF experiments using bremsstrahlung, the transitions in this energy range are quite difficult to separate from the sizable atomic background. It is highly desirable to remeasure the Dy isotopes by performing NRF experiments using quasi-monochromatic beams in the interesting energy region between 2 and 4 MeV as done for $^{232}$Th (see Ref.\cite{Adekola2011}). 

The integration limits for the SR function are given in the footnotes of Table~\ref{tab:SR}.
The table also includes the nuclear deformation
$\delta \approx \beta_2\sqrt{45/16\pi}$, where $\beta_2$ is
taken from Raman \textit{et al.}~\cite{raman2001} 
as compiled in Ref.~\cite{RIPL2}.

The observation that the $B_{\rm SR}$ strength in the Oslo-type of experiments using 
NRF integration limits compares well with the NRF results, is very interesting. 
It indicates that an eventual different effective moment of inertia for the SR built 
on the ground state compared to the SR embedded in the quasi-continuum has minor influence on
the strength. These findings are therefore fully consistent with the generalized Brink hypothesis.

\subsection{Radiative neutron capture cross sections}
Using our data as input, we have performed calculations of the radiative neutron capture, $(n,\gamma)$,
cross sections for the $^{159-163}\rm{Dy}$ target nuclei with the  reaction
code TALYS-1.8~\cite{TALYS,koning12}. The radiative neutron capture rates depend strongly
on the $\gamma$SF, 
which we now provide based on the new experimental data. It is a great
advantage that we have data also below the neutron separation threshold; if not,
the shape and absolute strength of the modeled $\gamma$SF below the threshold would be much more uncertain.

We use the GLO modeled strength functions with the experimental parameter
values found in Tables~\ref{tab:GDR_param} and ~\ref{tab:SR} for
the GEDR, PDR and SR structures.
The radiative neutron capture cross sections also rely on the level densities,
which we describe with the CT model fitted to our data and normalized to the experimental $s$-wave spacing $D_0$ values and the HFB calculations as described previously.
In addition, the cross sections depend on the neutron optical model potential (n-OMP),
for which we have applied the phenomenological approach by Koning and Delaroche~\cite{koning03}.
Also, we have used both the default option for the width-fluctuation treatment (the Moldauer expression~\cite{moldauer,moldauer2}) and the Hofmann-Richert-Tepel-Weidenm\"{u}ller model~\cite{hrtw,hrtw2,hrtw3}.

\begin{figure*}[t]
 \begin{center}
 \includegraphics[clip,width=2.\columnwidth]{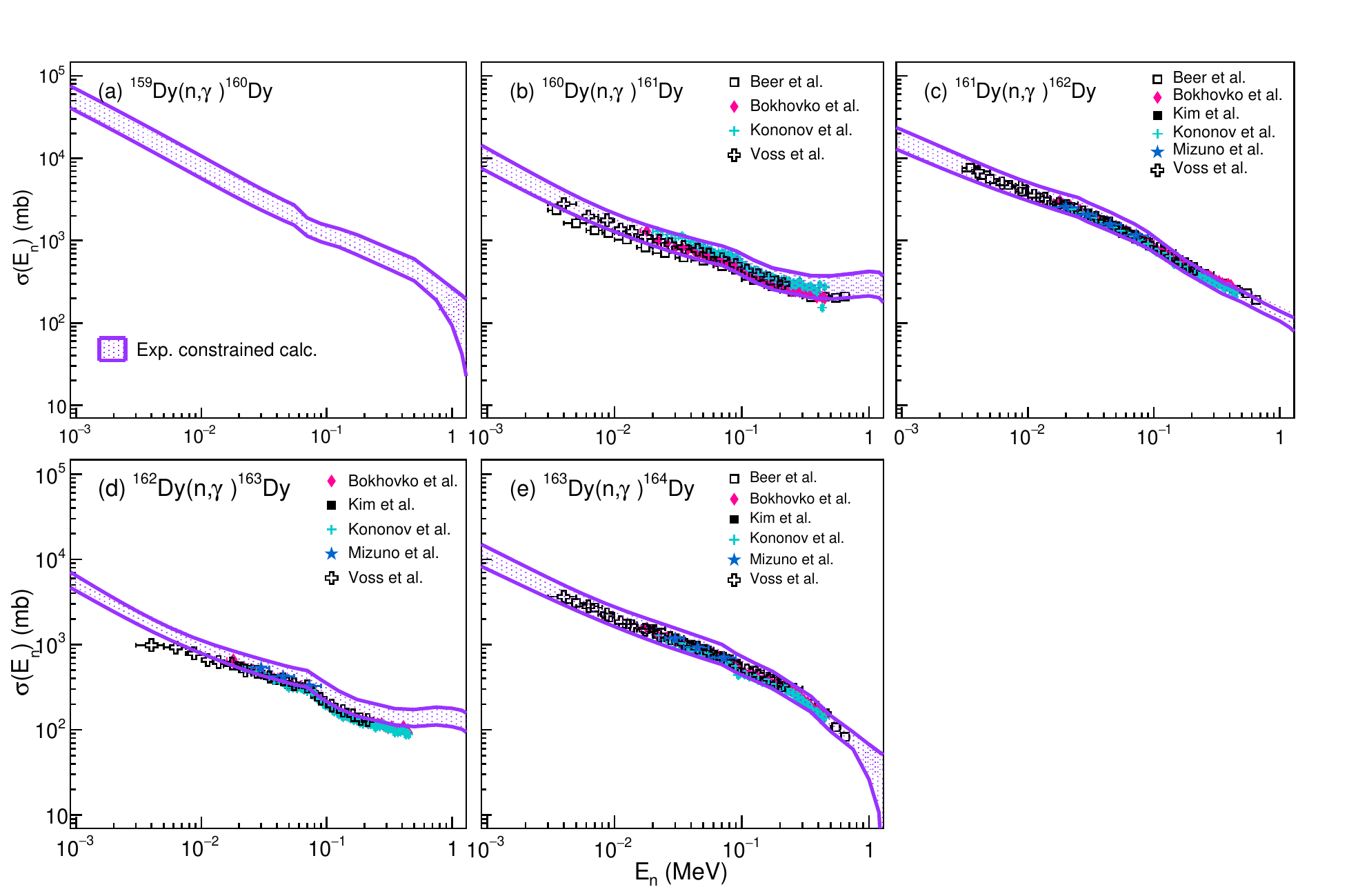} 
 \caption {(Color online) Calculated ($n,\gamma$) cross sections for 
$^{160-164}\rm{Dy}$ compound nuclei with TALYS-1.8~\cite{TALYS,koning12} compared
to data from Beer \textit{et al.}~\cite{beer1984}, Bokhovko \textit{et al.}~\cite{bohhovko1988}, Kim~\cite{kim2004}, Kononov \textit{et al.}~\cite{Kononov1981}, Mizuno \textit{et al.}~\cite{mizuno1999} and Voss \textit{et al.}~\cite{voss1999}. The purple, shaded areas display the calculated cross
sections using our data as input (included uncertainty of one standard deviation).}
 \label{fig:rsf_and_cross}
 \end{center}
 \end{figure*}

The resulting ($n,\gamma$) cross sections are compared to experimental
data  from Beer \textit{et al.}~\cite{beer1984}, Bokhovko \textit{et al.}~\cite{bohhovko1988}, Kim~\cite{kim2004}, Kononov \textit{et al.}~\cite{Kononov1981}, Mizuno \textit{et al.}~\cite{mizuno1999} and Voss \textit{et al.}~\cite{voss1999} as seen in Fig.~\ref{fig:rsf_and_cross}.
To our knowledge, for the case of $^{159}$Dy, this is the first time an experimentally-constrained 
($n,\gamma$) cross section is derived. 
For the other cases, where data exists, there is an excellent agreement with measured $(n,\gamma)$ cross sections,
within the experimental uncertainties. This fact gives further support to the applicability of detailed balance and the generalized Brink hypothesis for Dy isotopes.

\section{Summary and conclusions}  
\label{sec_slutt}
Using laser Compton backscattered $\gamma$-ray beams, we have extracted photo-neutron cross
sections of $^{162,163}\rm{Dy}$ above $S_n$. 
Gamma-ray strength functions above $S_n$ are deduced from the measured cross sections and are compared to reanalyzed data of $^{160-164}\rm{Dy}$ in the energy range below $S_n$. 
We observe that the $\gamma$SFs from the two different experimental approaches match well in both absolute value and slope at $S_n$ for $^{162,163}$Dy. This verifies the principle of detailed balance for absorption and emission of $\gamma$ rays with energies in the region of $E_{\gamma}\approx S_n$. 

By a careful determination of the underlying $E1$ component of the $\gamma$SF, we have evaluated the SR parameters for $^{160-164}\rm{Dy}$. 
All SR parameters agree well with each other and
average values of $\omega_{\rm SR} = 2.77(10)$ MeV and $B_{\rm SR} = 5.1(20) \mu_N^2$ are found.
The inclusion of several improvements in the analysis gives slightly different results than previously reported. 
However, we believe the new values are better founded. 

Based on the new photo-neutron data and the $^3\rm{He}$ induced reactions we
have calculated the inverse neutron capture cross sections using the reaction code TALYS.
The simulated radiative neutron capture cross sections are compared to experimentally
measured cross sections, and are in excellent agreement with the measurements.

In this work, the uncertainties quoted are carefully estimated including statistical as well as systematic errors. 
The summed strengths are compared to NRF and (n,$\gamma$) measurements on Dy isotopes. 
Provided that we use integration limits for the summed SR strength similar to ones used for NRF experiments, we find the same strength.  
The present results, therefore, confirm the validity of the generalized Brink hypothesis for the SR and the applicability of detailed balance for $\gamma$ decay and absorption in the $^{160-164}\rm{Dy}$ isotopes.

\acknowledgments 
T.R. and H.T.N. gratefully acknowledge financial support from the Research Council of Norway (NFR), project number 210007.
A. C. L. gratefully acknowledges funding from the ERC-STG-2014 under grant agreement no. 637686.
This work was supported by the Japan Private School Promotion Foundation and partly by
the JSPS-FNRS bilateral program. D.M.F., O.T. and I.G. acknowledge the support from the Extreme Light
Infrastructure Nuclear Physics (ELI-NP) Phase II, a project co-financed by
the Romanian Government and the European Union through the European
Regional Development Fund - the Competitiveness Operational Programme
(1/07.07.2016, COP, ID 1334).

\end{document}